\documentclass[prl,twocolumn,amsmath,mathtools,a4paper,showpacs]{revtex4-1}
\pdfoutput=1
\usepackage{graphicx}
\expandafter\let\csname equation*\endcsname\relax
\expandafter\let\csname endequation*\endcsname\relax
\usepackage{amsmath}
\usepackage{bm}
\usepackage{epsfig}
\usepackage{charter}
\usepackage{amssymb}
\usepackage{setspace}
\usepackage{csquotes}
\usepackage{float}
\usepackage{braket}
\usepackage{epstopdf}
\usepackage{xspace}

\begin{document}
	
\title{A.C.\ susceptibility as a probe of low-frequency magnetic dynamics}
\author{C.\ V.\ Topping\footnote{Present address: School of Physics and Astronomy, University of St Andrews, North Haugh, St Andrews, Fife KY16 9SS, United Kingdom} and S.\ J.\ Blundell}
\address{University of Oxford, Department of Physics, Clarendon Laboratory, Parks Road, Oxford OX1 3PU, United Kingdom}
\date{\today}

\begin{abstract}
The experimental technique of a.c.\ susceptibility can be used as a
probe of magnetic dynamics in a wide variety of systems.  Its use is
restricted to the low-frequency regime and thus is sensitive to
relatively slow processes.  Rather than measuring the dynamics of
single spins, a.c.\ susceptibility can be used to probe the dynamics
of collective objects, such as domain walls in ferromagnets or vortex
matter in superconductors.  In some frustrated systems, such as spin
glasses, the complex interactions lead to substantial spectral weight
of fluctuations in the low-frequency regime, and thus
a.c.\ susceptibility can play a unique role.  We review the theory
underlying the technique and magnetic dynamics more generally and give
applications of a.c.\ susceptibility to a wide variety of experimental
situations.

\vspace{0.5cm}

\noindent
{\bf Please cite as:} \newline 
C. V. Topping and  S. J. Blundell\newline
J. Phys.: Condens. Matter {\bf 31}, 013001 (2019)

\vspace{0.5cm}
\noindent
\url{https://doi.org/10.1088/1361-648X/aaed96}
\end{abstract}

\maketitle

\section{Introduction}\label{Intro}

The measurement of d.c.\ magnetic susceptibility is commonly used to characterise a newly discovered magnetic material.  Such a measurement can allow the elucidation of various magnetic properties of materials such as the presence of a phase transition, the magnetic moment of a material or simply the sign of magnetic exchange. However, in a d.c.\ measurement, the assumption is made that the sample properties remain effectively static, so that there is no measurable dynamic response.  This assumption can be restated in  terms of the dynamics being much faster, or slower, than the experimental timescale.  While this assumption holds in many cases, there are many classes of material where it does not.  In such cases, much useful information can be gained by employing a.c.\ magnetic susceptibility, a technique which utilises a periodic magnetic field rather than a static d.c.\ magnetic field.  
A.c.\ magnetic susceptibility has found application in various areas such as molecular magnetism \cite{GatteschiBook,Balanda2013}, ferromagnetism \cite{Chen1996} and superconductivity \cite{Nikolo1994,Gomory1997} and may be used to help differentiate between different types of slow relaxation \cite{GatteschiBook,Balanda2003,Banerjee2005, MydoshBook} and derive energy barriers for that relaxation \cite{GatteschiBook, Balanda2013, Chen1996}.

In this paper, we aim to provide a unified description of a.c.\ susceptibility.  We first contrast the a.c.\ and d.c.\ techniques in section~\ref{sec:acdc} and then, in section~\ref{sec:theory}, introduce a.c.\ susceptibility within the framework of linear response theory, highlighting similarities and differences with dielectric relaxation.  We outline methods of modelling real a.c.\ susceptibility data and illustrate these approaches with their applications to various classes of material in section~\ref{sec:applications}.

\section{D.C.\ and A.C.\ Magnetic Susceptibility}
\label{sec:acdc}
Magnetic susceptibility, $\chi$, is defined by the equation
\begin{equation}
\chi=\lim\limits_{H\rightarrow 0}\frac{M}{H},\label{GenSusceptEq}
\end{equation}
where $M$ is the sample magnetisation and $H$ is the applied magnetic field.  It is sometimes defined as a differential susceptibility
\begin{equation}
\chi=\frac{\partial M}{\partial H}.\label{Eq:LinChiDef}
\end{equation}
In the (commonly encountered) cases when $\chi\ll 1$, $B=\mu_0(H+M)\simeq\mu_0H$, and so
\begin{equation}
\chi \simeq \lim\limits_{B\rightarrow 0} \frac{\mu_0M}{B}.
\end{equation}
These equations apply to both d.c.\ and a.c\ magnetic susceptibility.  However, in a real experiment for the signal to be measurable we require the applied field to be of sufficient strength to produce that signal, and so the limit of vanishing field cannot be achieved.  Since $M$ generally is not linear with $H$
and susceptibility can be field dependent, one can define the differential
susceptibility
\begin{equation}
\chi_{\rm exp}=\frac{\delta M}{\delta H}\simeq \mu_0\frac{\delta M}{\delta B},
\end{equation}
where $\delta B=\mu_0 \delta H$ is a finite applied magnetic field.

\subsection{D.C.\ Susceptibility}
For a d.c.\ measurement, $\delta B$ is a small, static d.c.\ magnetic field (typically in the range 0.001--0.1\,T, though values outside this can be used) and the resulting magnetization $\delta M$ is recorded.  Typically, d.c.\ magnetic susceptibility of a system is measured as a function of temperature in two separate warming cycles. First, the sample will be cooled in zero applied magnetic field before applying the measurement field $\delta B$ and measuring $\delta M(T)$ at a number of fixed temperatures on the first warming cycle.  This is the zero field cooled (ZFC) sweep and the data recorded in this sweep probe the system taken out of steady state conditions.  Second, the sample is re-cooled but this time with the measurement field $\delta B$ and the warming cycle of measurements is repeated.  This yields the field cooled (FC) sweep and the data recorded correspond to the system in the steady state.  Thus, measurement of FC and ZFC sweeps can give an indication of the presence of slow magnetic relaxation, but the relevant timescale that is being probed depends on the rate at which both the magnet can be swept and a measurement can be made (if the dynamics are faster than this, the ZFC and FC sweeps will be identical). Nevertheless, d.c.\ susceptibility  remains a powerful tool for material study. 

Another d.c.\ technique that can study slow dynamics involves the direct measurement of any slow magnetic relaxation following the sudden removal of a magnetic field which has been applied for some time to a sample held at a constant temperature \cite{slow-relaxation}.  This can be referred to as d.c.\ relaxation, remanence or $M(t)$ dependence and is useful when the relaxation time is several tens of seconds, or even several hours, but is difficult to measure when the relaxation time is shorter than time needed to remove the magnetic field.

\subsection{A.C.\ Magnetic Susceptibility}\label{ACSect}
In a.c.\ magnetic susceptibility, a time varying, sinusoidal magnetic field of amplitude $H_{\rm a.c.}$ (typically $\sim 0.5$\,mT, though other values can be used) is applied to the sample.  Simultaneously, a static, d.c.\ magnetic field ($H_{\rm d.c.}$) may also be applied, though often this is set to zero (and the a.c.\ measurement is then directly probing the ground state of the spin system due to the small a.c.\ amplitude).  Thus the field $H$ inside
the sample is given by
\begin{equation}\label{acField}
H=H_{\rm d.c.}+H_{\rm a.c.}\cos(\omega t),
\end{equation}
where $\omega$ ($=2\pi\nu$) is the frequency of the oscillating magnetic field.  The frequency $\nu$ is typically in the range $0.1$--$10^4$\,Hz and so probes processes which are faster than those studied by the magnetic relaxation technique described above.  In this paper we will restrict our discussion to the situation in which the a.c.\ and d.c.\ magnetic fields are applied in parallel.  The oscillating response of the magnetisation is recorded ($M_{\rm a.c.}$) and the a.c.\ susceptibility is then defined by 
\begin{equation}
\chi_{\rm a.c.} = \frac{M_{\rm a.c.}}{H_{\rm a.c.}}.
\end{equation}
This equation has assumed that the response of the system is linear so that $M_{\rm a.c.}$ is proportional to $H_{\rm a.c.}$ with the constant of proportionality $\chi_{\rm a.c.}$; this is however not always the case and we will consider such nonlinear response in section~\ref{sec:nonlinear}.

\section{Theory and models}
\label{sec:theory}
In this section we will consider the theoretical background to measurements of a.c.\ susceptibility.  Before considering linear response in section~\ref{sec:linearresponse}, we will in section~\ref{sec:threetimes} provide a physical motivation for the relationship between the frequency $\omega$ and the characteristic relaxation time $\tau$ of the system.  In sections~\ref{sec:dampedSHO}--\ref{sec:rangetimes} we will explore various models of a.c.\ susceptibility, as well as ways of representing the response in the complex plane in section~\ref{sec:colecole}.  Non-linear effects will be considered in section~\ref{sec:nonlinear} and then in section~\ref{sec:susceptometer} we will describe the workings of a practical susceptometer.
\subsection{Three characteristic regimes}
\label{sec:threetimes}
Depending on the 
relaxation time $\tau$ of the magnetic moments of the studied system, three regimes can be defined on the basis of the relative sizes of $\omega$ and $1/\tau$.

\begin{figure} [tbp]
	\includegraphics[width = 1\columnwidth]{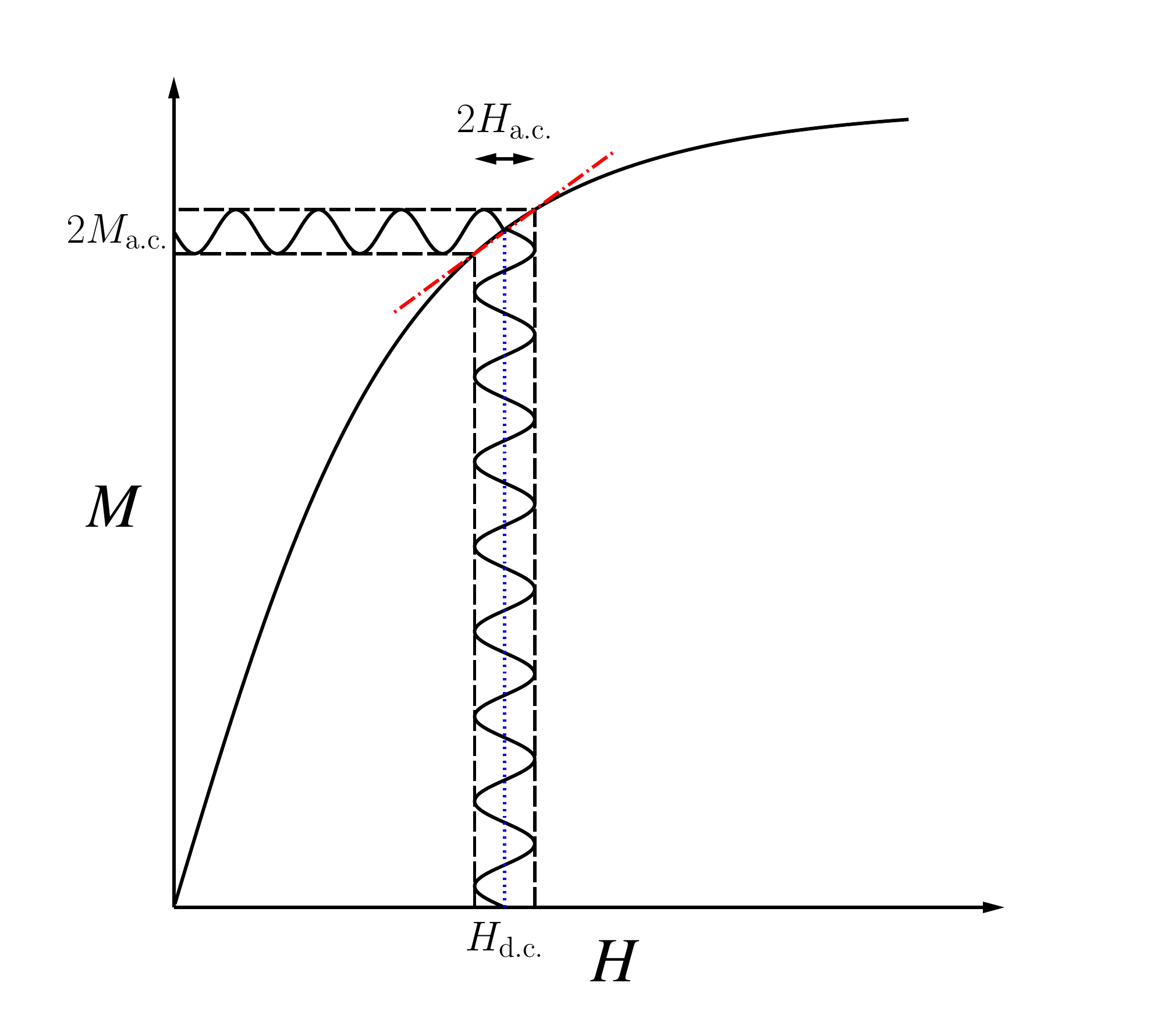}%
	\caption{Graphical demonstration of the use of a.c.\ magnetic susceptibility for measurement of the gradient of a magnetisation curve.  The red dashed line demonstrates the gradient being measured and the blue dotted line shows the applied d.c.\ magnetic field which can be changed to allow different parts of the magnetisation curve to be investigated.\label{fig:MvH_AC}}
\end{figure}

\begin{figure} [tbp]
	\includegraphics[width = 1\columnwidth]{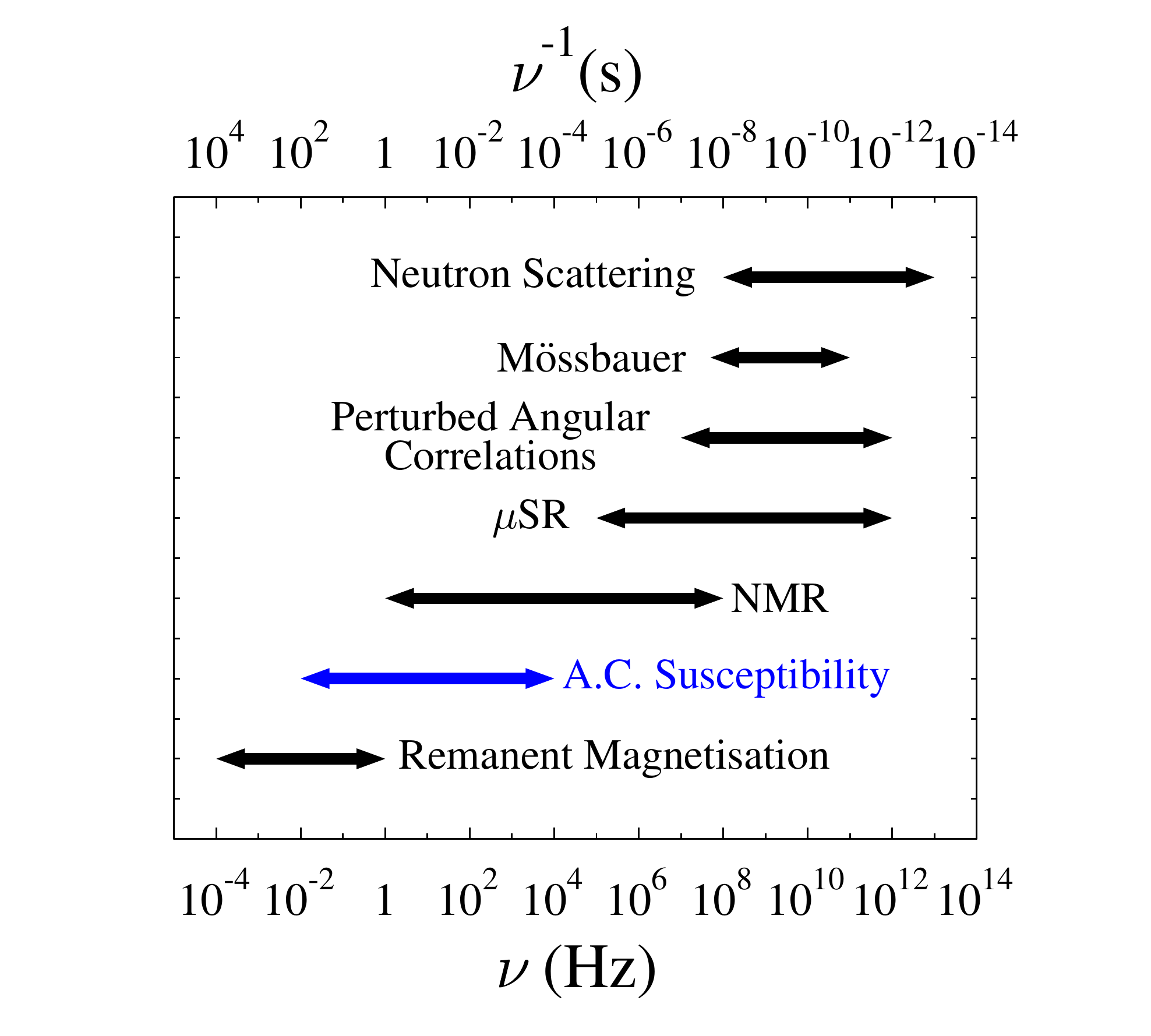}%
	\caption{Comparison of approximate frequency ranges available to various experimental techniques.  Information gathered from references~\cite{PPMSManual,YaouancBook}.  Note that (following equation~(\ref{RealImComp})) $\chi^{\prime\prime}$ has a maximum when $\omega\tau=1$ and so this occurs at $\tau=\nu^{-1}/2\pi$ and the factor of $1/2\pi$ should not be forgotten.\label{FreqRange}}
\end{figure}

(1) $\omega\ll 1/\tau$:   The first of these regimes corresponds to the d.c.\ limit in which the studied system responds essentially instantaneously to the a.c.\ field and d.c.\ susceptibility is obtained ($\chi_{\rm a.c.} \approx \chi_{\rm d.c.}$).  This is an equilibrium response and the moments are able to exchange energy with the lattice.  This results in a measurement of what we can call the isothermal susceptibility, $\chi_{T}$ \cite{GatteschiBook}.  Given the increased sensitivity that may be achieved due to measuring an oscillating response, a.c.\ susceptibility may be useful in studying systems where dynamics are not being considered but the signal is weak.  Furthermore, when using susceptibility as a measurement of the gradient of $M$ versus $H$, the ability to apply a d.c.\ magnetic field allows different regions of the $M$ versus $H$ curve to be probed as shown in figure~\ref{fig:MvH_AC}.

(2) $\omega\gg 1/\tau$: The second regime occurs when the perturbing field oscillates too quickly for the magnetic moments of the system to respond.  Thus the system does not have time to equilibrate and exchange energy with the lattice.  The obtained susceptibility is known as the adiabatic susceptibility, $\chi_{S}$ \cite{GatteschiBook}.

(3) $\omega\approx 1/\tau$: 
The intermediate regime, in which the frequency of the oscillating magnetic field is comparable to the timescale of the magnetic relaxation of the system, offers a much more complex response.  In this regime there may be some phase lag (and therefore dissipation) when the perturbation is slightly faster or slower than the natural frequency of the system.  Thus, the response is reported in two parts: in-phase and out-of-phase (or real and imaginary) components respectively, $M^{\prime}_{\rm a.c.}$ and $M^{\prime\prime}_{\rm a.c.}$, with corresponding susceptibility, $\chi^{\prime}_{\rm a.c.}$ and $\chi^{\prime\prime}_{\rm a.c.}$.  As shown below, the imaginary component relates to dissipation in the system. The a.c.\ magnetic susceptibility may be written as a complex number
\begin{equation}
\chi_{\rm a.c.} = \chi^{\prime}_{\rm a.c.} + i\chi^{\prime\prime}_{\rm a.c.},
\label{ACDef}
\end{equation}
which at low and high frequency must reduce to the real value ($\chi^{\prime\prime}_{\rm a.c.}=0$) of $\chi_T$ or $\chi_S$ respectively.  For brevity, the subscript ``a.c.'' will be neglected from this point onward.  The choice of the sign of the imaginary part of the susceptibility in equation~(\ref{ACDef}) is a matter of convention. In some treatments, the complex susceptibility is defined instead as $\chi^\prime-i\chi^{\prime\prime}$, and we will switch to this alternative choice later in this paper (in Section~\ref{sec:debye}). 

Ideally, when choosing a technique to examine dynamic behaviour frequencies that allow the probing of all three regimes should be available.  Figure~\ref{FreqRange} shows a comparison of several experimental techniques demonstrating both that a.c.\ magnetic susceptibility probes the lower frequency region and overlap does exist between techniques which can be taken advantage of should a system's characteristic time be at the edge of the available frequency region.

Descriptions of slow magnetic relaxation typically use a form known as the generalised Debye model. This has found application in various systems including single-molecule magnets \cite{GatteschiBook}, spin glasses \cite{MydoshBook} and ferromagnets \cite{Chen1996}.  This model was originally derived and applied to dielectric systems \cite{Jonscher1981,Cole1941,DebyeBook}, though also arose in treatments of magnetic materials \cite{Gorter1936,Casimir1938}. Our approach will be to use linear response theory, outlined in  Section~\ref{sec:linearresponse}, and to show how it may lead to the generalised Debye model and other related models in the sections that follow.

\subsection{Linear response theory}
\label{sec:linearresponse}
An arbitrary system will show a generalised displacement, $x(t)$, as a result of a generalised force, $f(t)$.  The value of $x$ at time $t$ is then given by

\begin{equation}
x(t)=\int_{-\infty}^{\infty}\chi(t-t^\prime)f(t^\prime){\rm d}t^\prime,
\end{equation}
where $\chi(t-t^\prime)$ is a generalised response function.  This relation is a convolution and hence we can write it as a product,

\begin{equation}
\tilde{x}(\omega)=\tilde{\chi}(\omega)\tilde{f}(\omega),
\end{equation}
by using the Fourier transform of $x(t)$, $\chi(t)$ and $f(t)$, explicitly defined as 

\begin{eqnarray}
\tilde{x}(\omega) &=& \int_{-\infty}^{\infty}e^{-i\omega t}x(t) {\rm d}t;\\
x(t) &=& \int_{-\infty}^{\infty}\frac{1}{2\pi}e^{i\omega t}\tilde{x}(\omega) {\rm d}\omega.
\end{eqnarray}
To simplify notation, we will drop the tildes on Fourier transforms and write them as $x(\omega)$, $f(\omega)$, $\chi(\omega)$, etc.  Furthermore, we will assume that $\chi(t-t^\prime) = 0$ for $t < t^\prime$ (an assumption of causality) and hence we can write $\chi(t) = X(t)\theta(t)$ where $\theta(t)$ is the Heaviside step function.
The function $X(t) = \chi(t)$ for $t>0$, but can take any value for $t<0$, so let us set it to $X(t) = -\chi(|t|)$ for $t<0$, making $X(t)$ an odd function (and hence $X(\omega)$ is purely imaginary).  Then
\begin{equation}
\chi(\omega) = \frac{1}{2\pi}\int_{-\infty}^{\infty}\theta(\omega^\prime - \omega)X(\omega){\rm d}\omega^\prime,
\end{equation}
and using $\theta(\omega)=\pi \delta(\omega)-i/\omega$, we have
\begin{equation}
\chi(\omega) = \frac{1}{2}X(\omega)-\frac{i}{2\pi}P\int_{-\infty}^{\infty}\frac{X(\omega^\prime)}{\omega^\prime-\omega}d\omega^\prime\\
= \chi^\prime(\omega)+i\chi^{\prime\prime}(\omega),
\label{eq:chiomegaprincipal}
\end{equation}
where $\chi^\prime$ and $\chi^{\prime\prime}$ are the real and imaginary parts of $\chi(\omega)$ and P indicates that the Cauchy principal value is taken to avoid a singularity.  Because $X(\omega)$ is purely imaginary, equation~\ref{eq:chiomegaprincipal} implies that
\begin{equation}
i\chi^{\prime\prime}(\omega) = \frac{1}{2}X(\omega)
\end{equation}
and
\begin{equation}
\chi^\prime(\omega) = P\int_{-\infty}^{\infty} {\rm d}\omega' \frac{\chi^{\prime\prime}(\omega^\prime)}{\pi(\omega^\prime-\omega)}.
\label{eq:kk}
\end{equation}
Equation~(\ref{eq:kk}) is one of the Kramers-Kronig relations which connects the real and imaginary parts of the response functions.
At $\omega=0$ equation~(\ref{eq:kk}) reduces to

\begin{equation}
\pi \chi^\prime(0) = P\int_{-\infty}^{\infty} {\rm d}\omega' \frac{\chi^{\prime\prime}(\omega^\prime)}{\omega^\prime}.
\label{eq:pichi}
\end{equation}
The quantity $\chi^\prime(0)$ is called the static susceptibility.

\subsection{The damped harmonic oscillator}
\label{sec:dampedSHO}
\begin{figure} [tbp]
	\includegraphics[width = 0.8\columnwidth]{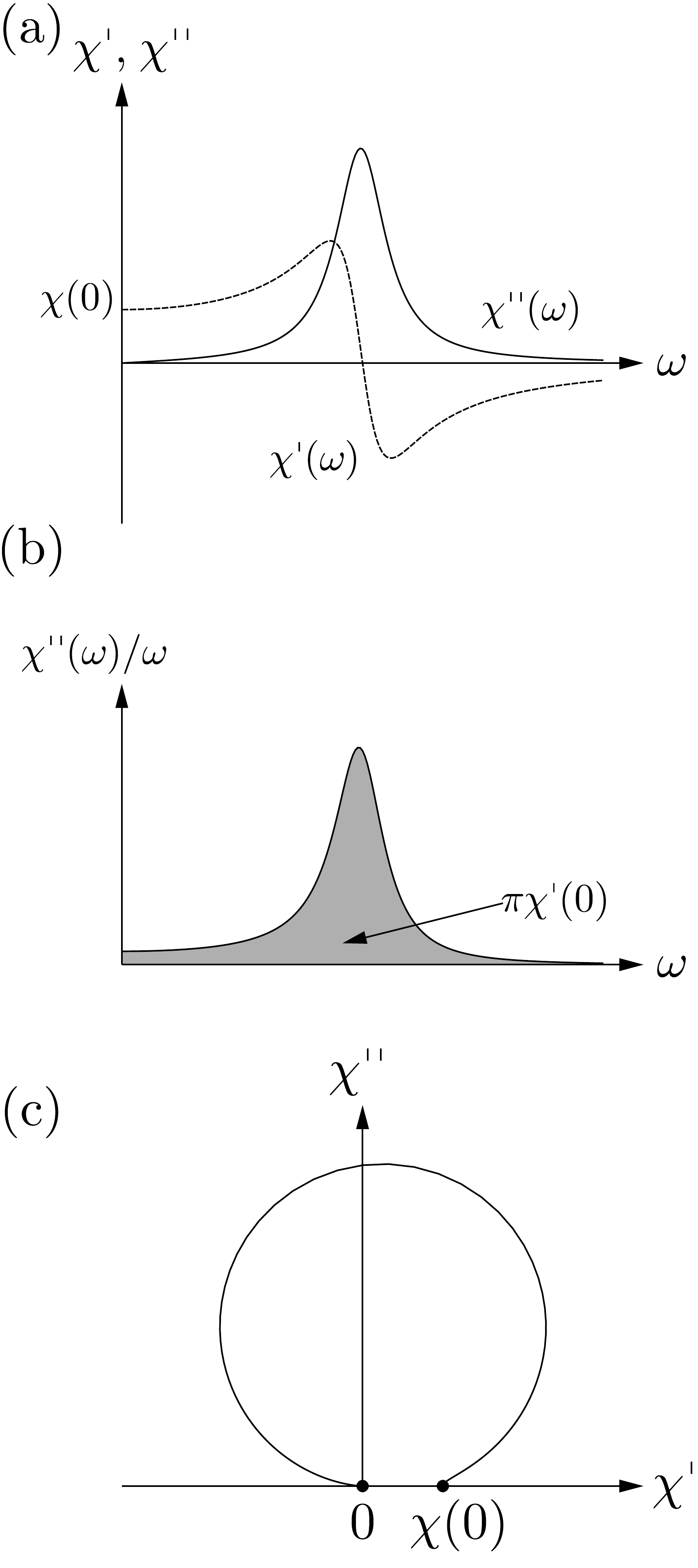}%
	
	\caption{(a) The real and imaginary parts of $\chi$ as a function of $\omega$. (b) Illustration of equation~(\ref{eq:pichi}) for the damped harmonic oscillator. (c) The same curves as in panel (a) but plotted in an Argand diagram. \label{chirealchiimag}} 
\end{figure}

A damped harmonic oscillator serves as an example of this approach.  The equation of motion is given by
\begin{equation}
m\ddot{x}+\alpha\dot{x}+kx=f,
\end{equation}
where $m$ is the mass, $\alpha$ is the damping constant and $k$ is the spring constant.  Writing the resonant frequency, $\omega_0^2=k/m$, and damping, $\gamma = \alpha/m$, we have
\begin{equation}
\chi(\omega)=\frac{x(\omega)}{f(\omega)}=\frac{1}{m}\left[\frac{1}{\omega_0^2-\omega^2-i\omega\gamma}\right].\label{Accel}
\end{equation}
This is a complex function and the real and imaginary parts are plotted in figure~\ref{chirealchiimag}(a).  The imaginary part $\chi^{\prime\prime}(\omega)$ is
given explicitly by
\begin{equation}
\chi^{\prime\prime}(\omega)=\frac{\omega\gamma/m}{(\omega^2-\omega_0^2)^2+(\omega\gamma)^2}.
\end{equation}
The static susceptibility is $\chi^\prime(0)=1/m\omega_0^2=1/k$, and straightforward integration shows that the sum-rule in equation~(\ref{eq:pichi}), $\int_{-\infty}^{\infty}\chi^{\prime\prime}(\omega)/\omega d\omega = \pi \chi^\prime(0)$, is satisfied; this relation is shown in figure~\ref{chirealchiimag}(b).  For later reference, we also show a plot of $\chi''$ against $\chi'$ in an Argand diagram in figure~\ref{chirealchiimag}(c).

\begin{figure} [tbp]
	\includegraphics[width = 0.8\columnwidth]{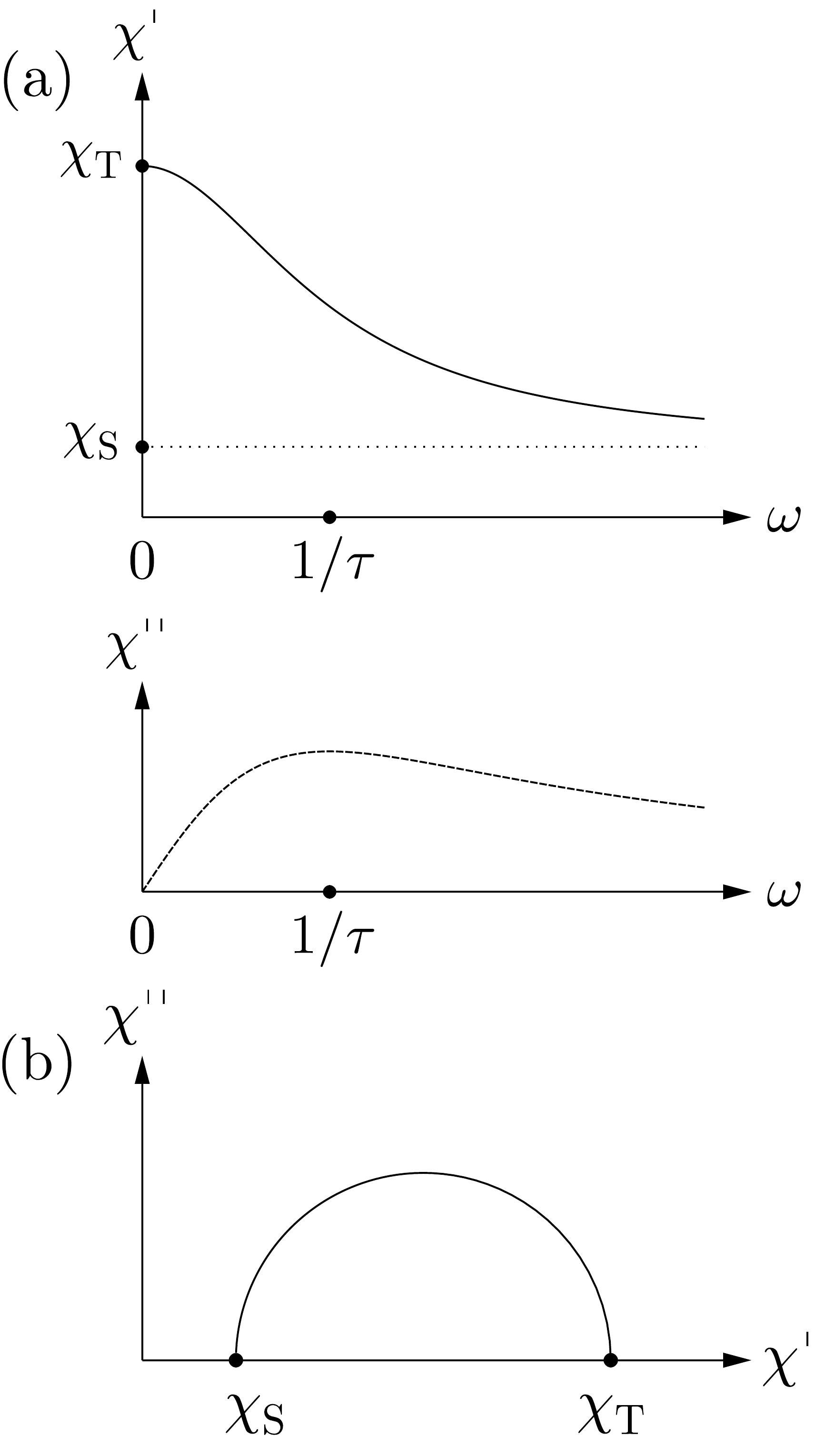}%
	
	\caption{(a) The real and imaginary parts of $\chi$ as a function of $\omega$ for the model with no inertial term.  The maximum in $\chi''$ occurs when $\omega\tau=1$, i.e.\ at $\omega=1/\tau$.  (b)  The same curves as in panel (a) but plotted in an Argand diagram (known as a Cole-Cole plot). \label{chirealchiimag2}} 
\end{figure}

Now we remove the inertial term resulting in the equation of motion becoming
\begin{equation}
\alpha\dot{x}+kx=f
\label{DifferentialDecay}
\end{equation}
with
\begin{equation}
\chi(\omega)=\frac{1/k}{1-i\omega\tau},
\end{equation}
where $\tau=\alpha/k$.  This yields real and imaginary parts

\begin{align}
  \chi^\prime(\omega)=\frac{1/k}{1+\omega^2\tau^2};\notag \\ \chi^{\prime\prime}(\omega)=\frac{\omega\tau/k}{1+\omega^2\tau^2}.  
\end{align}
Moreover, it is common to include an additive adiabatic response, $\chi_S$, so that
\begin{equation}
\chi(\omega)=\chi_S + \frac{1/k}{1-i\omega\tau}.\label{LinearDebye}
\end{equation}
Thus in the case of a.c.\ susceptibility, this expression becomes
\begin{align}
 \chi^\prime(\omega)=\chi_S+\frac{(\chi_T-\chi_S)}{1+\omega^2\tau^2};\notag \\ \chi^{\prime\prime}(\omega)=\frac{(\chi_T-\chi_S)}{1+\omega^2\tau^2}\omega\tau,\label{RealImComp}   
\end{align}
where we have written $\chi_T = \chi_S + 1/k$.  Here $\chi_S=\chi(\infty)$ is the adiabatic susceptibility and $\chi_T = \chi(0)$ is the isothermal susceptibility.  These equations recover the required limits at high and low frequencies, namely a reduction of $\chi^\prime$ to adiabatic and isothermal values respectively, as well as a vanishing of $\chi^{\prime\prime}$ at both limits.  These expressions are plotted in figure~\ref{chirealchiimag2}(a) as a function of $\omega$ (on a linear scale; to see these curves on a logarithmic scale of $\omega$, see the $\alpha=0$ curves in figure~\ref{DebyeModels}(a) and (d) below). Perhaps the most important feature present is the maximum of $\chi^{\prime\prime}$ which occurs at $\omega\tau = 1$ providing a convenient method to extract the relaxation time of a system.  In this model, the solution for equation~(\ref{DifferentialDecay}) with $f=0$ ($\tau \geq 0$) is given by
\begin{equation}
x(t)=x(0)e^{-{t}/{\tau}}
\end{equation}
and describes a relaxation process with a single relaxation time. 
For a magnetic system, $x(t)$ becomes $M(t)$, so that if a system has a magnetization which relaxes exponentially with time then it can be described by equation~(\ref{RealImComp}).
If $\chi''$ is plotted against $\chi'$ in the Argand diagram, a characteristic semicircular form is produced as shown in figure~\ref{chirealchiimag2}(b).

\subsection{Dielectric relaxation}
\label{sec:debye}
The model we have been considering (summarised in equation~(\ref{LinearDebye})) is analogous to the well-known expression for dielectric relaxation described in the Debye model \cite{DebyeBook}.  In Debye's treatment, the generalised displacement becomes the electric polarisation, $P(t)$, and the generalised force is an electric field, $E(t)$.  The model is usually formulated in terms of the relative permittivity $\epsilon(\omega) = 1 + P(\omega)/(\epsilon_0E(\omega))$.  Then, in the notation conventionally used to describe dielectric relaxation

\begin{eqnarray}
\epsilon^\prime = \epsilon_\infty+\frac{\epsilon_S-\epsilon_\infty}{1+\omega^2\tau^2};
\epsilon^{\prime\prime}=\frac{\omega\tau(\epsilon_S-\epsilon_\infty)}{1+\omega^2\tau^2},\label{ReIm}
\end{eqnarray}
where $\epsilon_S$ is the static permittivity ($= \epsilon_T$, the isothermal permittivity in the language we have adopted) and $\epsilon_\infty$ is the permittivity in the high-frequency limit ($= \epsilon_S$, the adiabatic permittivity in the language we have adopted).

Many of the treatments of a.c.\ susceptibility borrow expressions used in dielectric relaxation, and in the field of dielectric relaxation it is conventional to write the complex susceptibility in the Debye model as

\begin{equation}
\chi(\omega)=\chi_S + \frac{\chi_T-\chi_S}{1+i\omega\tau},\label{StatedDebye}
\end{equation}
where the sign difference in the denominator occurs due to the previously mentioned differing convention of complex susceptibility (which is defined as $\chi_{\rm a.c.}=\chi^\prime-i\chi^{\prime\prime}$ for the above equation).  We will use this convention from now on as it is the usual choice in the literature.  Both equations~(\ref{LinearDebye}) and (\ref{StatedDebye}) lead to the same real and imaginary parts of equations~(\ref{RealImComp}) (equivalent to the dielectric case given in equation~(\ref{ReIm})).  

The Debye model fails to model short times (and thus high frequencies) and violates the sum-rule that $\int_0^\infty \omega \chi''(\omega)\,{\rm d}\omega$ should remain finite, so modifications sometimes need to be considered if very high-frequency studies are carried out \cite{Onodera1993} which, for example, can be done using time-domain terahertz spectroscopy \cite{Pan2015}.  

\subsection{A range of relaxation times}
\label{sec:rangetimes}
The Debye model arises from linear response theory when one assumes that the variable of interest (whether  magnetisation or polarisation) relaxes according to a simple exponential relaxation.  Thus we assume that there is a slowly relaxing ``entity'' that relaxes with a single time scale.  This means that the entities cannot interact with each other because this can create clustering effects which lead to a distrubution of relaxation times.    Indeed it is an acknowledged problem in dielectric realxation that the ideal process predicted for non-interacting electric dipoles \cite{Jonscher1981,Cole1941,Mantas1999}  is rarely obtained \cite{Jonscher1981,Jonscher1977}.  Similarly, in the case of magnetism, the limit of completely non-interacting magnetic moments also seems unlikely to be encountered due to the presence of cooperative effects, though it might not necessarily be a bad approximation for systems such as superparamagnets \cite{SteveBook} or single-molecule magnets \cite{GatteschiBook}.

\begin{figure*} [tbp]
	\includegraphics[width = 2\columnwidth]{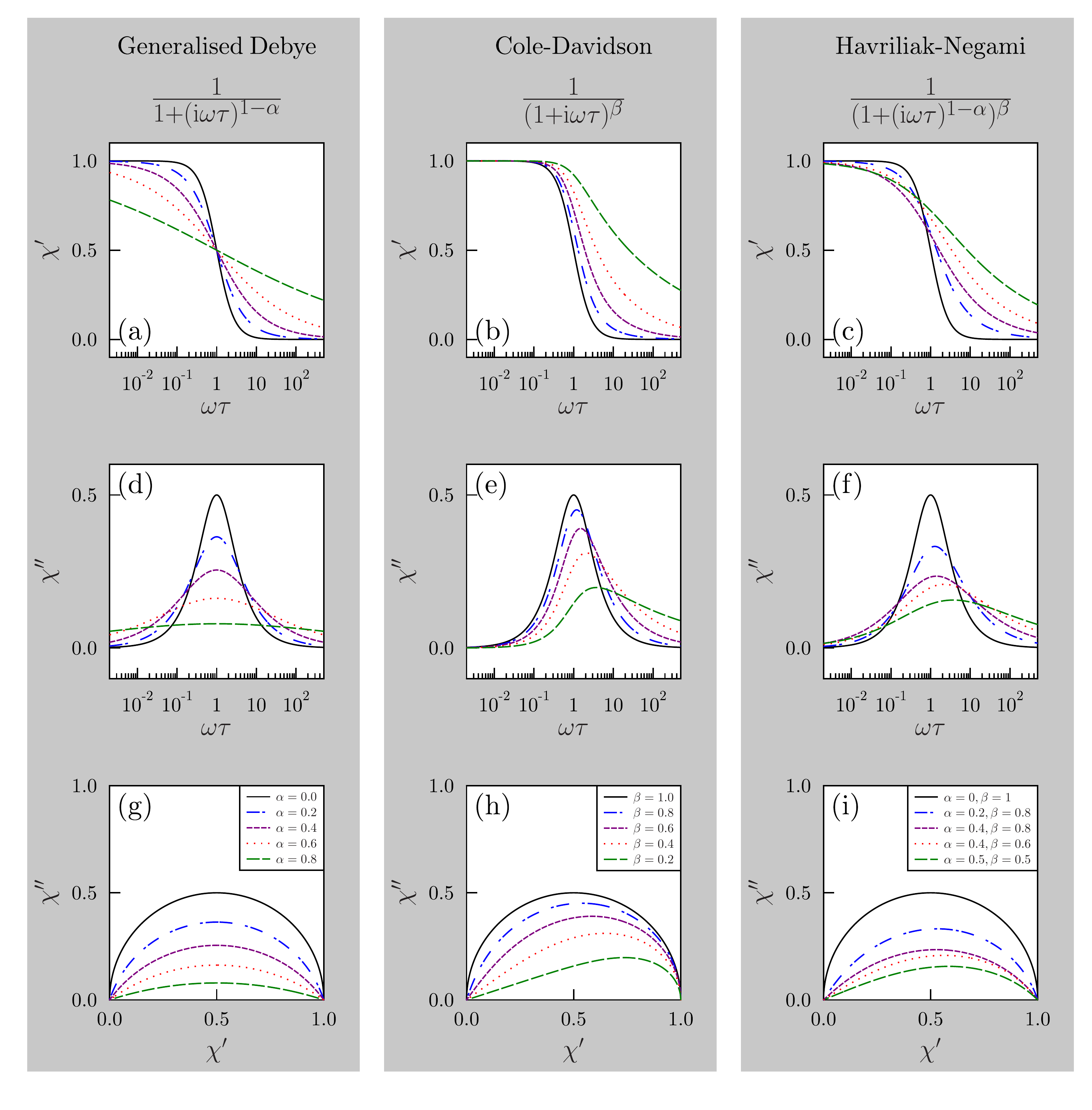}%
	\caption{Example real ((a)-(c)) and imaginary ((d)-(f)) parts of the a.c.\ response and the Cole-Cole plots ((g)-(i))  for the a.c.\ magnetic susceptibility interpretation of the Generalised Debye ((a),(d) and (g)), Cole-Davidson ((b), (e) and (h)) and Havriliak-Negami ((c), (f) and (i)) models.  These plots assume $\chi_{\rm S}=0$ and $\chi_{\rm T}=1$, though the scaling to general values is obvious.  The function $\chi(\omega)=\chi_{\rm S}+(\chi_{\rm T}-\chi_{\rm S})y(\omega)$, where $y(\omega)$ is the function shown at the head of each column in the figure. \label{DebyeModels}}
\end{figure*}

In order to account for these complexities, several approaches can be employed.  One strategy is to introduce a spread of relaxation times into the model.  This is often accounted for by introducing a phenomelogical parameter $\alpha$ into what is then called the generalised Debye model

\begin{equation}
\chi(\omega)=\chi_S+\frac{\chi_T-\chi_S}{1+(i\omega\tau)^{1-\alpha}},\label{GenDebye}
\end{equation}
where $0\leq\alpha\leq 1$ \cite{GatteschiBook,Cole1941}.  Setting $\alpha=0$ corresponds to no spread of relaxation times and the ideal Debye model is recovered.  This modification (shown in figures~\ref{DebyeModels}(a) and (d)) is successful in describing slowly relaxing electric \cite{Cole1941} and magnetic systems \cite{GatteschiBook,Balanda2013,Goura2015,Hagiwara1998}.  An alternative approach is known as the Cole-Davidson model (shown in figures~\ref{DebyeModels}(b) and (e)) \cite{Davidson1951} which instead places an exponent $\beta$ in the denominator as follows:

\begin{equation}
\chi(\omega)=\chi_S+\frac{\chi_T-\chi_S}{(1+i\omega\tau)^{\beta}}.\label{Cole_Davidson}
\end{equation}
In the study of dielectric systems these equations are sometimes combined to produce the Havriliak-Negami equation 

\begin{equation}
\chi(\omega)=\chi_S+\frac{\chi_T-\chi_S}{[1+(i\omega\tau)^{1-\alpha}]^{\beta}}\label{Havriliak-Negami},
\end{equation}
which (by virtue of introducing two variable parameters) can improve the agreement with data from real systems \cite{Havriliak1967,Havriliak1994}.  Plots of this model with various parameter values are shown in figures~\ref{DebyeModels}(c) and (f).  However, the introduction of additional fitting variables risks overparameterization.

\begin{figure} [tbp]
	\includegraphics[width = \columnwidth]{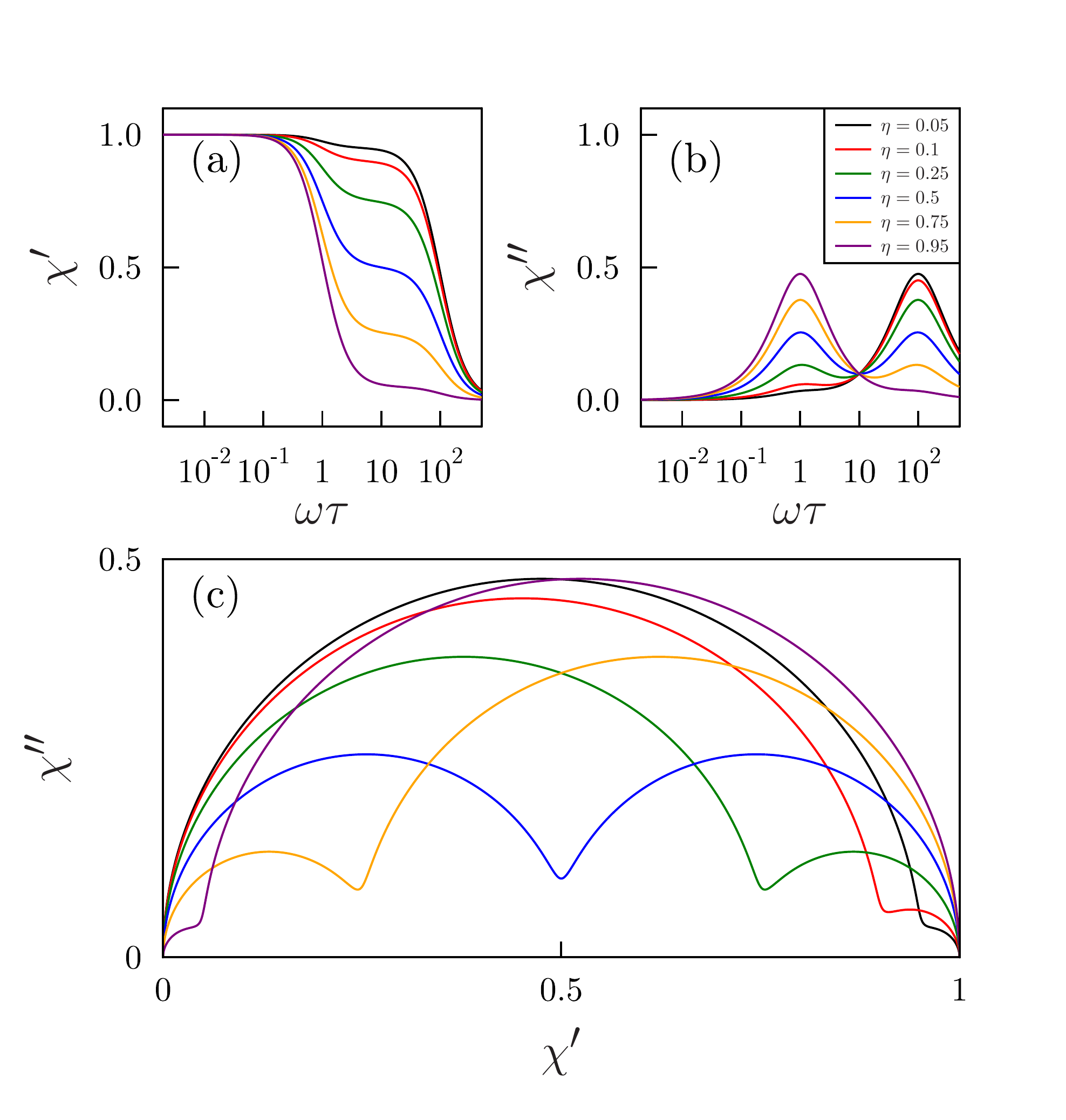}%
	\caption{The real (a) and imaginary (b) parts of $\chi$ and the Cole-Cole plot (c) for a model using equation~(\ref{eq:twomodel}) with $\tau_1/\tau_2=100$ ($\tau_1=\tau$, $\tau_2=0.01\tau$) and plotted for different values of $\eta$.  \label{TwoDebye}}
\end{figure}

Each of the above extensions of the Debye model (equations~(\ref{GenDebye})--(\ref{Havriliak-Negami})) are all somewhat ad hoc adjustments introduced to yield improved agreement between real systems and theory (for example, reference~\cite{Davidson1951} refers to the Cole-Davidson model as an empirical formula).  Therefore, their use which will be shown for single-molecule magnets, spin glasses and spin ices in Sections~\ref{sect:SMMs}, \ref{sect:SpinGlasses} and \ref{sect:SpinIce} respectively tends to arise because of the typical models employed in these areas.  Common to each of these extensions is the assumption that a single relaxation time no longer governs the system dynamics, with a distribution of relaxation times  parameterised by $\alpha$ and/or $\beta$, depending on the model used.  The effect of multiple relaxation times can be illustrated by considering a magnetic system composed of two magnetic entities with distinct relaxation times $\tau_1$ and $\tau_2$.  The total magnetisation of the system is then the sum of each entity's magnetisation, and so the susceptibilities will also add.  Hence

\begin{multline}
\chi = \chi_1 + \chi_2 = \\ \chi_{S,1} + \frac{\chi_{T,1}-\chi_{S,1}}{1+i\omega\tau_1} + \chi_{S,2} + \frac{\chi_{T,2}-\chi_{S,2}}{1+i\omega\tau_2}.\label{Simp2tau}
\end{multline}
If each entity has the same magnetic moment and $\eta$ is the fraction of the system composed of the first entity and $1-\eta$ the fraction composed of the second entity this reduces to

\begin{equation}
\chi = \chi_{S} + (\chi_{T}-\chi_{S})\left(\frac{\eta}{1+i\omega\tau_1} + \frac{1-\eta}{1+i\omega\tau_2}\right),\label{eq:twomodel}
\end{equation}
where we have assumed each entity relaxes exponentially in a Debye-like manner.  (A similar approach is used in
reference~\cite{Mantas1999}, with a different definition of constants.) 
This model gives two maxima in $\chi^{\prime\prime}$ at $\omega=1/\tau_1$ and $\omega=1/\tau_2$ which are easy to distinguish if there is a large enough separation between the two relaxation times.  This approach however retains the limitations intrinsic in the Debye model (such as the neglect of interactions). Some simulations based on the model of equation~(\ref{eq:twomodel}) (for which the two time constants are assumed to be separated by a factor of one hundred) are shown in figure~\ref{TwoDebye}. This demonstrates the two maxima in $\chi^{\prime\prime}$ and also shows how two arcs are generated in the Cole-Cole plot; Cole-Cole plots undergoing further discussion in the next section.  If $\tau_1$ and $\tau_2$ are not too dissimilar, this results in a single asymmetric arc in the Cole-Cole plot \cite{Mantas1999}.  This can look rather similar to the arc produced in the Cole-Davidson model (see figure~\ref{DebyeModels}(h)).

This approach can be extended to an arbitrary number of coexisting processes with different relaxation times.  For example,  equation~(\ref{eq:twomodel}) can be generalised to

\begin{equation}
\chi(\omega) = \chi_{S} + (\chi_{T}-\chi_{S})\sum\frac{\eta_n}{1+i\omega\tau_n},
\end{equation}
where $\eta_n$ is the proportion of the system with relaxation time $\tau_n$ and $\sum\eta_n=1$.  If the number of relaxation processes is large, their distribution can be replaced by a continuous function and the summation may be replaced by an integration

\begin{equation}
\chi(\omega) = \chi_{S} + (\chi_{T}-\chi_{S})\int_{\tau_{\rm min}}^{\tau_{\rm max}}\frac{g(\tau)}{1+i\omega\tau}d\tau,\label{DistribDebye}
\end{equation}  
with $g(\tau)$ as the distribution of relaxation times.

The precise form of $g(\tau)$ depends on the system in question.  In the case of a single relaxation time $\tau_{\rm c}$, $g(\tau) = \delta(\tau -\tau_{\rm c})$ and the ideal Debye model with $\tau=\tau_{\rm c}$ is recovered as applicable to systems such as superparamagnets \cite{SteveBook} and single-molecule magnets \cite{GatteschiBook}.  Various forms of distributions of relaxation times have been considered and are listed in reference~\cite{Zorn2002}, though in that work the distribution of relaxation times $f(\ln(\tau))$ is defined in terms of the logarithm of the relaxation time, so that the integral in equation~(\ref{DistribDebye}) is written as
$\int {\rm d}\ln\tau\,f(\ln\tau)/(1+i\omega\tau)$ (though the two forms can easily be related using ${\rm d}\ln\tau=\tau^{-1}\,{\rm d}\tau$, so that $\tau^{-1}f(\ln\tau)=g(\tau)$).

\begin{figure*}[tbp]
	\includegraphics[width=2\columnwidth]{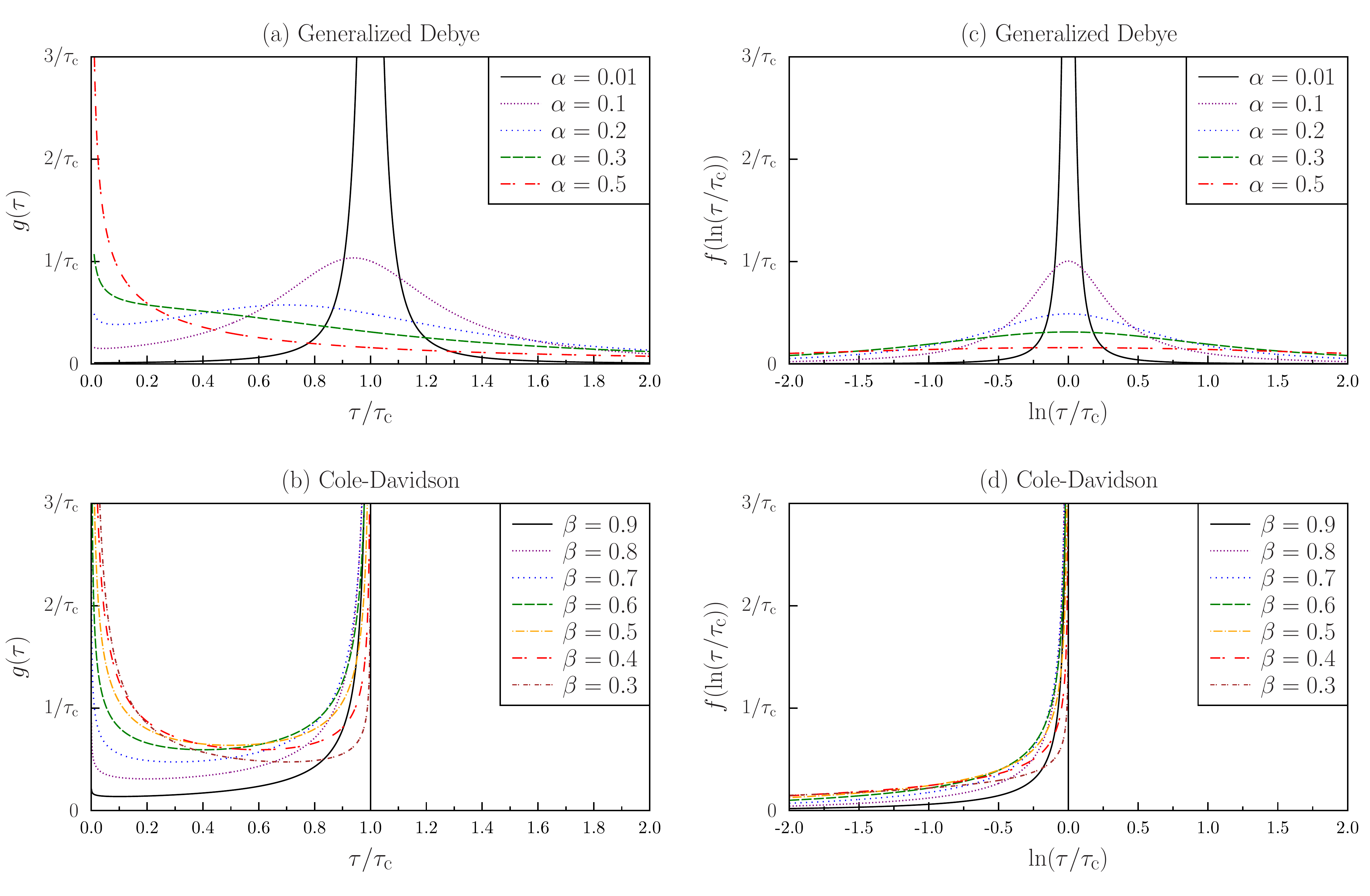}%
	\caption{The left-hand panels show the form of $g(\tau/\tau_{\rm c})$ in (a) the generalized Debye model (plotted for different values of $\alpha$) and (b) the Cole-Davidson model (plotted for different values of $\beta$). 
	The right-hand panels show the form of $f(\ln(\tau/\tau_{\rm c}))$ for different values for (c) the generalized Debye model and (d) the Cole-Davidson model.  Thus these are analogous to the plots in panels (a) and (b), but plotted as a function of $\ln(\tau/\tau_{\rm c})$.
	\label{fig:coledav}}
\end{figure*}

An example of this approach is shown in
figure~\ref{fig:coledav}(a) which contains plots of the distribution function of the generalised Debye model for different values of $\alpha$.  The formula for this distribution function \cite{Huser1986} is
\begin{equation}
    g(\tau) = \frac{1}{2\pi\tau} \frac{\sin\alpha\pi}{\cosh[(1-\alpha)\ln(\frac{\tau}{\tau_{\rm c}})]-\cos\alpha\pi}.\label{eq:DistriGenDebye}
\end{equation}
As $\alpha$ approaches zero, $g(\tau)$ gets more sharply peaked near $\tau=\tau_{\rm c}$, becoming $g(\tau) = \delta(\tau -\tau_{\rm c})$ for $\alpha=0$ (the ideal Debye model limit).  For larger values of $\alpha$ one finds the distribution to be broader.

The same can be done for the Cole-Davidson form
$\chi(\omega)=\chi_S+(\chi_T-\chi_S)/[(1+i\omega\tau_{\rm c})^{\beta}]$ (given earlier in equation~(\ref{Cole_Davidson}), but note that here we are writing the characteristic time as $\tau_{\rm c}$).  To do this, 
following \cite{Davidson1951} we choose the form of $g(\tau)$ in equation~(\ref{DistribDebye}) as
\begin{equation}
    g(\tau) = 
    \begin{cases} 
    \displaystyle{\sin{\beta\pi}\over\pi} {1 \over \tau} \left( {\tau \over \tau_{\rm c} - \tau } \right)^\beta & \tau\leq \tau_{\rm c} \\
    0       & \tau > \tau_{\rm c}.
  \end{cases} \label{eq:DistriCD}
\end{equation}
This function is plotted in figure~\ref{fig:coledav}(b) for various values of $\beta$.  As $\beta\to 1$, the function becomes more strongly peaked at $\tau=\tau_{\rm c}$ (and becomes a delta function at $\beta=1$, the ideal Debye model limit).  For smaller values of $\beta$ there is a broader distribution of $\tau$. 
In contrast with the generalised Debye model, the distribution has an upper cut-off in $\tau$. 

The analogous expressions for the Havriliak-Negami form
$\chi(\omega)=\chi_S+(\chi_T-\chi_S)/([1+(i\omega\tau_{\rm c})^{1-\alpha}]^{\beta})$ (given earlier in equation~(\ref{Havriliak-Negami})) is \cite{Havriliak1967,Zorn1999,alvarez1991}
\begin{equation}
    g(\tau)={1 \over \pi\tau} 
    { \left( \frac{\tau}{\tau_{\rm c}}\right)^{(1-\alpha)\beta}\sin\beta\Theta \over 
    \left[ \left( \frac{\tau}{\tau_{\rm c}}\right)^{2(1-\alpha)} - 2\left( \frac{\tau}{\tau_{\rm c}}\right)^{1-\alpha}\cos\pi\alpha + 1 \right]^{\beta/2} } ,  \label{eq:DistriHN}
\end{equation}
where
\begin{equation}
    \Theta = \tan^{-1} \left| {\sin\pi\alpha \over \left( \frac{\tau}{\tau_{\rm c}}\right)^{1-\alpha} - \cos\pi\alpha }  
    \right| .
\end{equation}
Equation~(\ref{eq:DistriHN}) reduces to equation~(\ref{eq:DistriGenDebye})
when $\beta=1$ and to equation~(\ref{eq:DistriCD}) when $\alpha=0$.

We note that though these different expressions for $g(\tau)$ reproduce the various forms of $\chi(\omega)$, none of them has an obvious physical basis.  
One model providing a better fit to experimental data over another can suggest features of the actual distribution function of relaxation times that might be present.  For example, a good fit to the Cole-Davidson model might suggest the presence of an upper cut-off in the relaxation time distribution with a long tail, rather than a distribution that is smeared out on either side of $\tau_{\rm c}$).  However, a distribution of relaxation times could arise from interactions between the relaxing entities and describing this in detail for a real system is a complicated problem, outside the scope of these phenomenological models.  In principle, some other contribution of processes could account for the experimental data just as well.   The plots in figure~\ref{fig:coledav}(a) and (b) are replotted in figure~\ref{fig:coledav}(c) and (d), but using a logarithmic time axis (and writing $\tau^{-1}f(\ln(\tau/\tau_{\rm c}))=g(\tau)$).

Sometimes it is possible to make some statements about a distribution of relaxation times that have a better motivated physical basis.
A useful relation that has been applied to spin glasses can be derived
by assuming that the distribution of relaxation times is very broad
and relatively uniform over several decades, i.e.\ over
a wide range of $\ln\tau$; thus let us assume that $f(\ln\tau)=\bar{f}$ 
(where $\bar{f}$ is a constant)
between $\tau_{\rm min}$ and $\tau_{\rm max}$, and further that
$\omega$ lies somewhere in the middle of this range, so that
$\tau_{\rm min}\ll \omega^{-1} \ll \tau_{\rm max}$.  In this case,
using equation~(\ref{DistribDebye}), $\chi'$ can be written
\begin{equation}
  \chi'=  \chi_{S} + (\chi_{\rm T}-\chi_{\rm S})\int_{\tau_{\rm
      min}}^{\tau_{\rm max}}\frac{f(\tau)}{1+\omega^2\tau^2}d\ln\tau,
\end{equation}
and the gradient of $\chi'$ as a function of $\ln\omega$ is
\begin{eqnarray}
  {\partial\chi' \over \partial\ln\omega}
  &=& \omega{\partial\chi' \over \partial\omega} \\
  &=& -2 (\chi_{\rm T}-\chi_{\rm S}) \bar{f} \int_{\tau_{\rm
      min}}^{\tau_{\rm
      max}}\frac{\omega^2\tau^2}{1+\omega^2\tau^2}d\ln\tau \\
  &=& (\chi_{\rm T}-\chi_{\rm S}) \bar{f} \left[
    \frac{1}{1+\omega^2\tau_{\rm max}^2}
    -           \frac{1}{1+\omega^2\tau_{\rm min}^2}
     \right] \\
  &\approx& -(\chi_{\rm T}-\chi_{\rm S}) \bar{f} ,
  \label{eq:lundgren1}
\end{eqnarray}
where the last approximation follows from using the limits in the form
$\omega\tau_{\rm min}\ll 1$ and $\omega\tau_{\rm max}\gg 1$.
Similarly, $\chi''$ can be written
\begin{eqnarray}
    \chi''&=&  (\chi_{\rm T}-\chi_{\rm S})\int_{\tau_{\rm
      min}}^{\tau_{\rm max}}\frac{f(\tau)
      \omega\tau}{1+\omega^2\tau^2}d\ln\tau, \\
    &=& (\chi_{\rm T}-\chi_{\rm S}) \bar{f} [\tan^{-1}(\omega\tau_{\rm
      max})
    - \tan^{-1}(\omega\tau_{\rm min})] \\
    &\approx& -(\chi_{\rm T}-\chi_{\rm S}) \bar{f} ,
  \label{eq:lundgren2}    
\end{eqnarray}
Comparing equations~(\ref{eq:lundgren1}) and (\ref{eq:lundgren2})
yields
\begin{equation}
  \chi'' \approx {\pi\over 2}   {\partial\chi' \over \partial\ln\omega},
\label{eq:lundgren3}
\end{equation}
a relationship between $\chi'$ and $\chi''$, first derived by Lundgren
{\sl et al.} \cite{Lundgren1981}, that holds quite well in various spin glass
systems \cite{Lundgren1982}.

\subsection{The Cole-Cole plot}
\label{sec:colecole}
In order to differentiate between different models and determine which best describe a studied system it is helpful to look at a Cole-Cole (or Argand) plot of $\chi^{\prime\prime}$ versus $\chi^\prime$ \cite{GatteschiBook,Cole1941}.  For an ideal Debye process ($\alpha=0$ in the generalised Debye model) such a plot shows a semicircle with the flat side on the $x$-axis (as has already been introduced in figure~\ref{chirealchiimag2}(b)).  When a spread of relaxation times is introduced as for the generalised Debye model the semicircle becomes distorted and sinks below the $x$-axis, the arc angle made with the $x$-axis being $(1-\alpha)\pi$ (see Appendix~A).  An example of this is shown in figure~\ref{DebyeModels}(g) with figures~\ref{DebyeModels}(h) and (i) showing a comparable Cole-Cole plot for the Cole-Davidson and Havriliak-Negami models.  Thus for example, if data can be described by the generalised Debye model then the temperature dependence of $\alpha$ can be extracted by considering a set of these arcs at various temperatures and fitting them to the following relation between  $\chi^{\prime\prime}$ and $\chi^{\prime}$: 

\begin{multline}
  \chi'' = - \left(\frac{\chi_{\rm T}-\chi_{\rm S}}{2}\tan\frac{\pi\alpha}{2}\right)
  \\ \pm
  \sqrt{
   \left(\frac{\chi_{\rm T}-\chi_{\rm
       S}}{2}\tan\frac{\pi\alpha}{2}\right)^2
   + (\chi'-\chi_{\rm S})(\chi_{\rm T}-\chi')
  }.
\label{ColeColeTest}  
\end{multline}
Here $\alpha$ is a fitting parameter (see, for example,~\cite{Hagiwara1998}).  Equation~(\ref{ColeColeTest}) can be derived from equation~(\ref{GenDebye}), as shown in Appendix~A.
(If a dataset does not extend over a sufficiently large frequency range to carry out this kind of fit, it is still possible to extract $\alpha$ using the angle the data makes with the $x$-axis in the Cole-Cole plot \cite{Cole1941}). 
In contrast the Cole-Davidson model (equation~(\ref{Cole_Davidson})) shows a non-symmetric arc in a Cole-Cole plot with one side elongated \cite{Davidson1951} (see figure~\ref{DebyeModels}(h), as well as Appendix~B).

In an a.c.\ susceptibility experiment, the shape of the arcs of a Cole-Cole plot (symmetric for the generalised Debye model and asymmetric for the Cole-Davidson and Havriliak-Negami models) tend to define which model is employed.  As such, no particular physical reasoning is usually employed when selecting a model for a system beyond the appearance of the Cole-Cole plot, and so this remains a phenomenological model.

In an ideal situation, an
a.c.\ susceptibility experiment would involve the measurement of $\chi^{\prime}$ and $\chi^{\prime\prime}$ over a large frequency range so that the condition $\omega\tau = 1$ is satisfied for every significant relaxation process.   In practice, data over a small set of particular frequencies are obtained as temperature is varied due to equipment constraints.  If the relaxation time of the system varies with temperature ($\tau(T)$), then for a certain frequency the condition $\omega\tau(T) = 1$ may be satisfied at some temperature \cite{GatteschiBook}.  Assuming $\chi_T$ and $\chi_S$ vary slowly in this temperature region, a peak will occur in a plot of $\chi^{\prime\prime}$ vs $T$ when this condition is satisfied, for a particular $\omega$, thereby allowing $\tau(T)$ to be modelled   \cite{GatteschiBook,Balanda2013,Binder1986,Novikov2015,Boca2017}, a point we will return to in Section~\ref{sect:SMMs}.


\subsection{Nonlinear effects and harmonics}
\label{sec:nonlinear}
So far we have considered only linear response, motivated by the definition in equation~\ref{Eq:LinChiDef} that states that $M$ and $H$ are linearly related by $\chi$.  However, in general $M$ might be a more complicated function of $H$, so in that case we can write  $M$ as a polynomial expansion in  $H$ \cite{Banerjee2005,MydoshBook,Cimberle2006,Fujiki1981,Bitoh1996}:
\begin{equation}
    M=M_0+\chi^{(1)}H+\chi^{(2)}H^2+\chi^{(3)}H^3+\ldots\label{eq:MPolyExpan}
\end{equation}
Here $M_0$ is a spontaneous magnetisation (which is zero in several of the systems considered in this paper) and $ \chi^{(1)}$ is the linear susceptibility that we have been discussing thus far.  If the applied magnetic field $H$ is small, then the nonlinear terms can be neglected, but sometimes they need to be considered.  Very often only odd powers of $H$ are required in equation~(\ref{eq:MPolyExpan}) due to the symmetry of $M$ \cite{Cimberle2006,Bitoh1996,Ozcelik1992}, but we will leave them all in.
With an applied field given by
\begin{equation}
    H=H_{\rm a.c.}\cos(\omega t)\label{eq:ACField}
\end{equation}
the resultant magnetisation $M(t)$ can be expanded as a Fourier series \cite{MydoshBook,Ishida1990}
    
\begin{equation}
    M=M_0 +\sum_{n=1}^\infty \mathcal{M}_n\cos(n\omega t).\label{Eq:FourierSeriesHarmonics}
\end{equation}
In this case, and assuming that $\chi^{(n)}$ are all real quantities, the form of equation~(\ref{eq:MPolyExpan}) yields the following expressions for the Fourier components:
\begin{eqnarray}
  \mathcal{M}_1 &=& \chi^{(1)}H_{\rm a.c.}+\frac{3}{4}\chi^{(3)}H_{\rm a.c.}^3
  + \frac{5}{8}\chi^{(5)}H_{\rm a.c.}^5 + \cdots \nonumber \\
  \mathcal{M}_2 &=& \frac{1}{2}\chi^{(2)}H_{\rm a.c.}^2+\frac{1}{2}\chi^{(4)}H_{\rm a.c.}^4
   + \cdots \nonumber \\
    \mathcal{M}_3 &=& \frac{1}{4}\chi^{(3)} H_{\rm a.c.}^3
  + \frac{5}{16}\chi^{(5)}H_{\rm a.c.}^5 + \cdots \nonumber \\
    \mathcal{M}_4 &=& \frac{1}{8}\chi^{(4)}H_{\rm a.c.}^4
  + \frac{3}{16}\chi^{(6)}H_{\rm a.c.}^6 + \cdots \nonumber \\
    \mathcal{M}_5 &=& \frac{3}{4}\chi^{(5)}H_{\rm a.c.}^5
   + \cdots .
\end{eqnarray}
The values of these Fourier components will come out differently if the
a.c.\ magnetic field in equation~(\ref{eq:ACField}) is chosen as a sine, rather than a cosine function  (see for example reference~\cite{Ozcelik1992}). Moreover, more complicated expressions can be derived if a d.c.\ magnetic field is also applied \cite{Ozcelik1992}.
Measuring the nonlinear susceptibility has been used to differentiate
between different types of slow magnetic relaxation \cite{Bitoh1996} due to the divergence in $\chi^{(3)}$ near the critical temperature \cite{MydoshBook,Fujiki1981,Suzuki1977} and can be very sensitive to the presence of some magnetic phases which are undetected in linear a.c.\ susceptibility \cite{Cimberle2006,Bitoh1993,Mito2015,Clements2018,Tsurata2018}.

A different treatment of nonlinear effects is used in studies of superconductors, as will be described in section~\ref{sec:superconductor}.  There a particular focus is on the  real {\it and} imaginary parts of the nonlinear susceptibility and explicit forms of the very nonlinear $M(H)$ behaviour according to different models of superconducting behaviour (rather than simply using a series expansion as in equation~(\ref{eq:MPolyExpan})) can be directly tested
 \cite{Ishida1990,Polichetti2012,Polichetti2004,Zola2004,Adesso2004,Adesso2004b}.

\subsection{A.C.\ Susceptometer}
\label{sec:susceptometer}

 \begin{figure*} [tbp]
 \centering
	\includegraphics[width = 2\columnwidth]{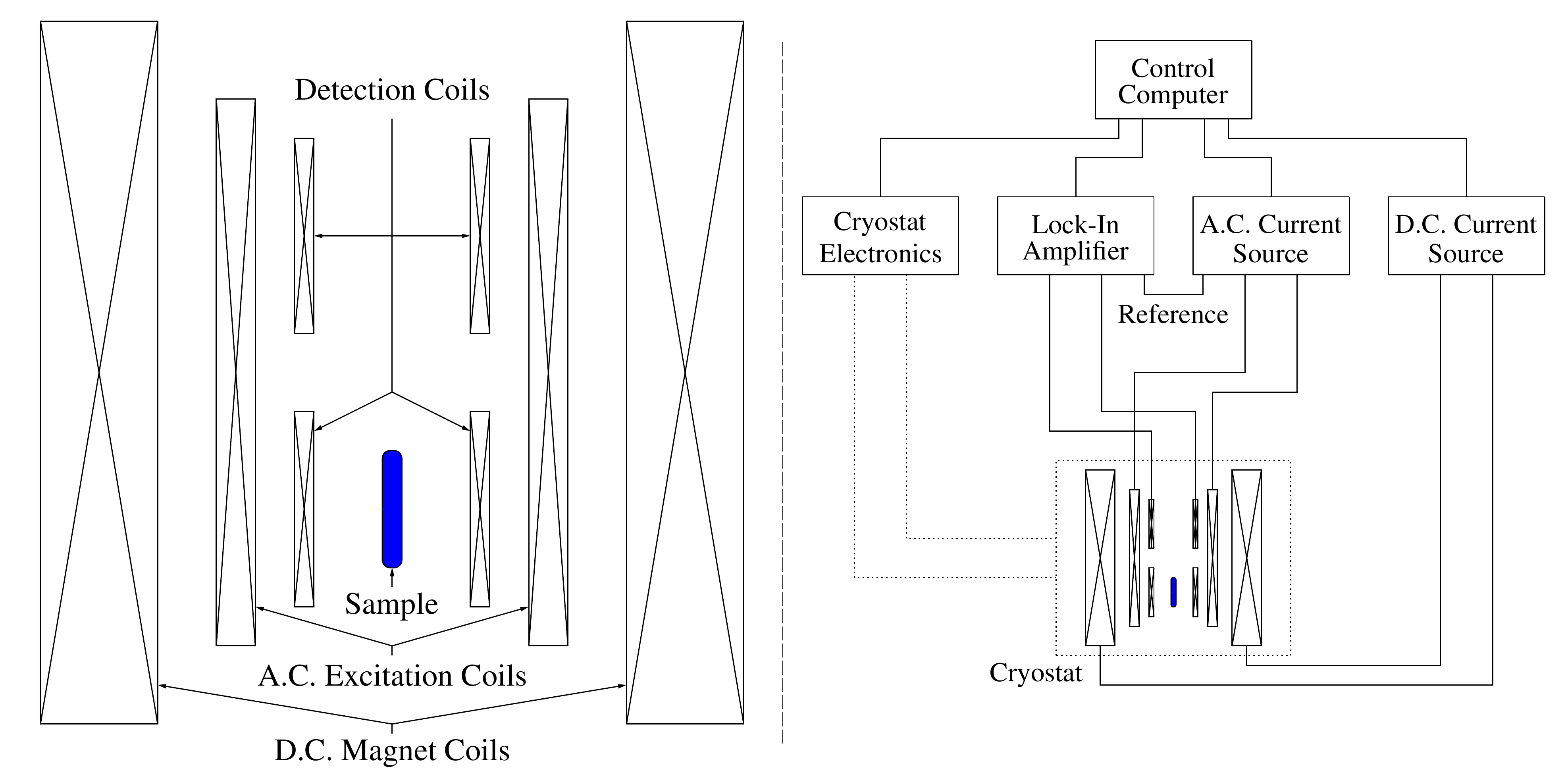}%
	\caption{Schematic of the physical apparatus of an a.c.\ susceptometer.  Left shows the set up of the various coils involved in the system.  Right shows the electrical connections in an a.c\ susceptometer.  \label{ACDiagram}}
\end{figure*}

Both a.c.\ susceptometry and d.c.\ susceptometry utilise the effect of a changing magnetic flux inducing a voltage in a detector/sensing coil (Faraday's law) and use the magnetisation of a sample to generate this changing flux \cite{Youssif2000}.  For a d.c.\ measurement a changing magnetic flux is achieved by physically translating the sample through the detection coil \cite{PPMSManual,Youssif2000}.  A.c.\ measurements generate a changing flux due to the applied a.c.\ magnetic field yielding a time-varying response in the sample with the sample kept stationary.

An a.c.\ susceptometer contains three  distinct coil sets: a.c.\ excitation coils, detector coils,  and d.c.\ magnet coils (see the left panel of figure~\ref{ACDiagram}).
The a.c.\ magnetic field is generated by an excitation coil set (sometimes called the primary or drive coils) which are driven by an a.c.\ current source providing the range of possible frequencies that may be accessed \cite{Nikolo1994,PPMSManual,Youssif2000,Edgar1993}.  The detector coils (sometimes called secondary coils) are typically placed within the a.c.\ excitation coils.  They consist of a pair of identical and connected oppositely wound coils with the sample located at the centre of one of these coils \cite{Nikolo1994,PPMSManual,Youssif2000}.  Setting up the system in this way with two detector coils of opposite handedness helps null signals originating form the a.c.\ field or other external sources by keeping one coil empty \cite{Nikolo1994}.  The signal is detected using a lock-in amplifier (taking a reference from the a.c.\ current source) thus allowing the in-phase ($\chi^\prime$) and out-of-phase ($\chi^{\prime\prime}$) components to be detected \cite{Nikolo1994,Edgar1993}.  Should higher harmonics be desired these can be detected using additional lock-in amplifiers at the appropriate multiples of the a.c.\ drive frequency.  This set-up is shown on the right of figure~\ref{ACDiagram}.

While the detector coil pair are nominally identical this is normally not perfectly achieved in practice requiring methods to compensate for incomplete nulling \cite{Nikolo1994}.  This can be accomplished by including a sample translation stage allowing measurements to be performed with the sample at different positions in the detector coil pair (such as in the centre of the lower detector coil as depicted in figure~\ref{ACDiagram}, in the centre of the upper detector coil or between the detector coil) \cite{PPMSManual}.  This is the method adopted by a Quantum Design Physical Property Measurement System using the AC Measurement System option \cite{PPMSManual}.  Accurate determination of the phase difference between the a.c.\ drive and sample signal is important (noting the in-phase sample signal is actually $\pi/2$ out-of-phase with the a.c.\ drive due to Faraday's law \cite{PPMSManual}).  Any additional phase differences introduced by the electronics must be accounted for \cite{PPMSManual}.  Moreover, the a.c.\ magnetic field can introduce heating problems at low temperature \cite{PPMSManual} and mechanical vibrations can affect measurement accuracy \cite{Couach1991}.

For samples with a large susceptibility, a demagnetization correction should be performed.  This is particularly important with a.c.\ measurements because failure to make an appropriate correction can lead to the real and the imaginary parts being mixed together.  Using SI units, the intrinsic susceptibility $\chi$ is related to the experimentally measured susceptibility $\chi_{\rm exp}$ by
$\chi^{-1} = \chi_{\rm exp}^{-1}-N$ where $N$ is the demagnetizing factor which depends on the shape of the sample \cite{SteveBook}.  Therefore the real and imaginary parts are
\begin{eqnarray}
\chi^{\prime} &=& 
{ \
\chi^{\prime}_{exp} - N ( [\chi^{\prime}_{exp}]^2 + [\chi^{\prime\prime}_{exp}] ) \over
(1-N \chi^{\prime}_{exp} )^2 + (N\chi^{\prime\prime}_{exp})^2
} \\
\chi^{\prime\prime} &=& 
{ \
\chi^{\prime\prime}_{exp}  \over
(1-N \chi^{\prime}_{exp} )^2 + (N\chi^{\prime\prime}_{exp})^2 
} .
\end{eqnarray}
These expressions must be used in studies on superconductors where $\chi^{\prime}_{\rm exp}$ and $\chi^{\prime\prime}_{\rm exp}$ is large.

\section{Application to real experimental systems}
\label{sec:applications}
The models outlined above can describe a wide range of slowly relaxing phenomena.    This section provides some examples of various families of experimental systems for which a.c.\ susceptibility is a useful tool as well as a discussion of the origin of the slow relaxation in each family.

\subsection{Paramagnetism}\label{Paramagents}

The first magnetic system to consider is paramagnetism.  In this state the magnetic moments are free to relax at a rate given by the spin-spin relaxation time which is very fast (i.e.\ $\approx 10^{-9}-10^{-10}~\rm s$) meaning that the system responds effectively instantaneously, at least on the timescales accessible to a.c.\ susceptibility \cite{GatteschiBook,Balanda2013,MorrishBook}.  Thus a paramagnet should not show slow relaxation.  In fact, $\chi_{\rm T}=\chi_{\rm S}$ in a paramagnet which sets $\chi^{\prime\prime}=0$ according to equation~(\ref{RealImComp}). The presence of a non-zero $\chi^{\prime\prime}$ can be indicative of a departure from paramagnetism.  
This statement is only true under
zero applied d.c.\ magnetic field.  The application of a d.c.\ field creates a net magnetization and the possibility of spin-lattice relaxation due to direct phonon processes, which are  accessible for study using a.c.\ susceptibility \cite{GatteschiBook}.

\subsection{Superparamagnets and Single Molecule Magnets}\label{sect:SMMs}
Superparamagnets and single molecule magnets (SMMs) are arguably the simplest systems exhibiting slow magnetic relaxation.  A superparamagnet describes an assembly of magnetic particles, each of which have sufficiently small physical size that domain wall formation is not possible and the magnetization in each particle becomes a single magnetic domain  \cite{GatteschiBook,SteveBook}  These are sometimes referred to as magnetic nanoparticles \cite{Gatteschi2012,Narayanasamy2018,Singh2009}.  SMMs (sometimes called zero-dimensional molecular magnets) are a subset of the greater class of molecular magnets \cite{GatteschiBook,Blundell2004}. In a SMM, each molecule contains magnetic ions which are linked by organic ligands (an example is shown in figure~\ref{fig:cadiou}(a) for the molecule Ni$_{12}$ which consists of twelve Ni$^{2+}$ ions and has a $S=12$ ground state).  The individual molecules are held together in a crystal only rather weakly and each molecule can therefore be considered to be a zero-dimensional magnet (the intermolecular exchange can largely be neglected) \cite{Blundell2004}.  The intramolecular exchange produces a net ``giant spin'' ground state (such as the $S=12$ ground state in Ni$_{12}$), so that each molecule can be considered as a single, giant magnetic moment to an approximation \cite{GatteschiBook,Gatteschi2003,Hill2010}.  Thus SMMs can show superparamagnetic behaviour.  The moments in both superparamagnets and SMMs are well isolated from each other and therefore are good candidates for exhibiting slow relaxation of the kind described by the Debye model. 

\begin{figure} [tbp]
 \fbox{\bf Copyrighted figure -- please see published article}
	\caption{(a) The single molecule magnet Ni$_{12}$ (chemical formula [Ni$_{12}$(chp)$_{12}$(O$_2$CMe)$_{12}$(H$_2$O)$_6$(THF)$_6$], where chp = 6-chloro- 2-pyridonate). (b) $\tau$ against $1/T$ measured from ac susceptibility (red circles) or d.c.\ relaxation (green triangles); inset: $\chi''(T)$ at 10 frequencies. Adapted from reference~\cite{Cadiou2001}.\label{fig:cadiou}}
\end{figure}

\begin{figure} [tbp]
	\includegraphics[width = 1\columnwidth]{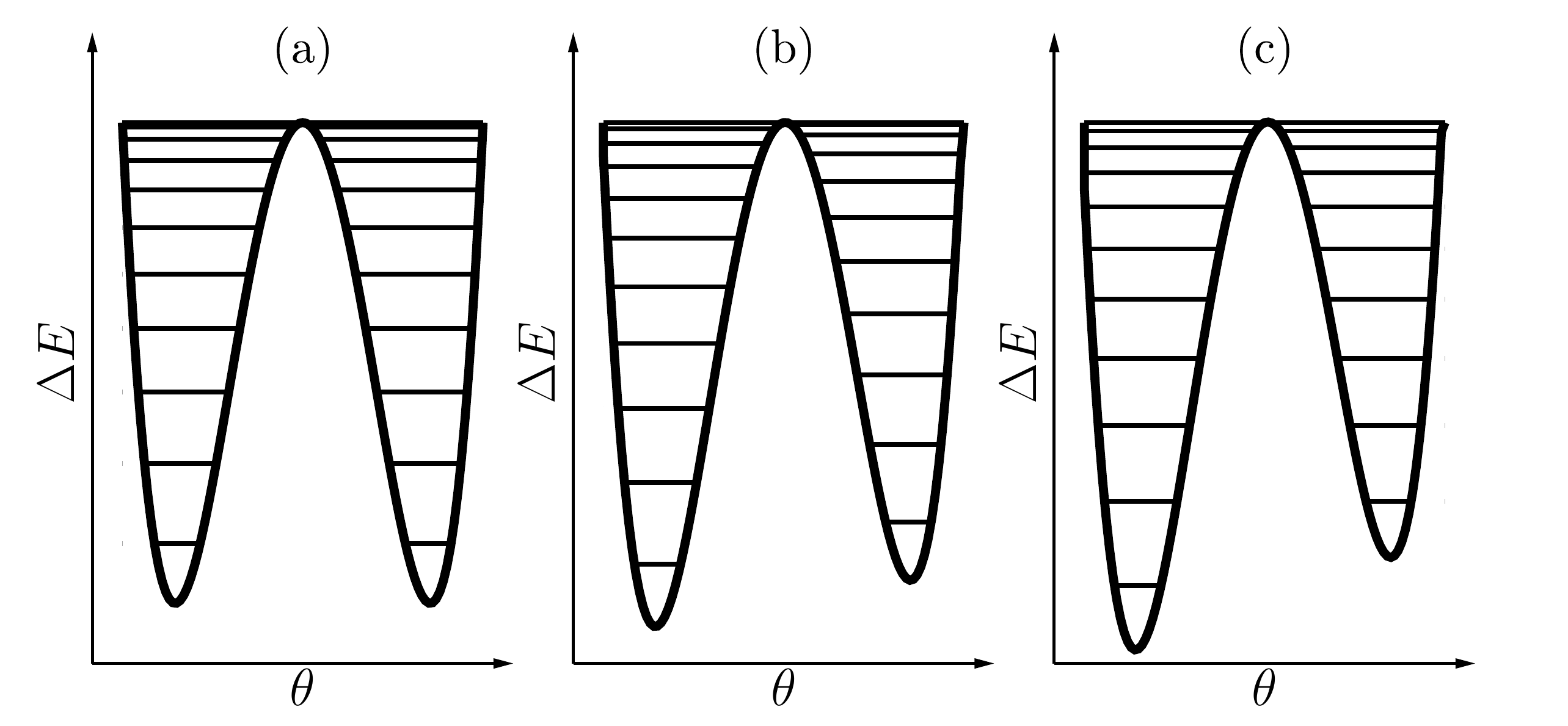}%
	
	\caption{Potential energy diagram for a system described by the Hamiltonian $H = -DS_z^2$ with an externally applied magnetic field of $B=0$ ((a)), $B = D/2g\mu_{\rm B}$ ((b)) and $B = D/g\mu_{\rm B}$ ((c)).  Adapted from reference~\cite{Blundell2004}.\label{DoubleWell}}
\end{figure}

The source of the slow magnetic relaxation in both systems arises from uniaxial anisotropy (i.e.\ they contain a magnetic easy axis) making it energetically favourable for moments to align along a specific axis.  For superparamagnets, this uniaxial anisotropy arises from magnetocrystalline anisotropy or shape anisotropy and can be described by an energy density ${\cal E}$ given by

\begin{equation}
{\cal E} = K\sin^2\theta,
\end{equation}
where $K$ is a constant describing the anisotropy energy density and $\theta$ is the angle made with the easy axis by the single domain moment \cite{SteveBook,Gatteschi2012}.  If $K>0$, energy is minimized when the moment is aligned parallel or anti-parallel with the easy axis.  Therefore, the potential energy diagram of the magnetic moment is a symmetric double well with an energy barrier separating the parallel and anti-parallel configurations.   For SMMs, the anisotropy results from a zero field splitting parameter, $D>0$, arising from a Hamiltonian $\mathcal{H}$ of the form

\begin{equation}
\mathcal{H} = -DS_z^2 + E(S_x^2 - S_y^2) + g \mu_{\rm B} \bm{B}\cdot\bm{S},
\label{SMMHamil}
\end{equation}
where $\bm{S}$ is the spin operator, $S_x$, $S_y$ and $S_z$ are the spin components along the $x$, $y$ and $z$-axes, $E$ is an additional (rhombic) anisotropy term,  $g$ is the g-factor and $\bm{B}$ is the magnetic field \cite{GatteschiBook,Blundell2004,Gatteschi2003,Hill2010}.  The term premultiplied by $E$ introduces a medium and hard axis and is essentially a higher order term that we will set to zero for our initial discussion\cite{GatteschiBook}.  The symmetric double well potential energy diagram this creates when $\bm{B}=0$ and $E=0$ (analogous to superparamagnets) is shown in figure~\ref{DoubleWell}(a).  The quantized $+S$ and $-S$ states lie on opposite sides of an energy barrier of height (or activation energy) $E_{\rm a}$.  (It should be noted that in some papers the parameter $D$ is defined with opposite sign).  
Slow magnetic relaxation arises from the moments overcoming this energy barrier to change orientation between parallel and anti-parallel states.  If this is a thermal process, the relaxation time $\tau$ is temperature-dependent and can be described using an Arrhenius law of the form

\begin{equation}
\tau(T) = \tau_0 \exp\left(\frac{E_{\rm a}}{k_BT}\right),
\label{eq:arrhenius}
\end{equation}
where $\tau_0$ is  known as the inverse attempt frequency (which can be thought of as the time between attempts at thermally exciting over the energy barrier) and can take values down to around $10^{-11}$\,s but can be many orders of magnitude larger \cite{GatteschiBook}.  $E_{\rm a}=KV$ for superparamagnets where $V$ is the volume of the single domain particle \cite{SteveBook}.  For SMMs, this type of behaviour is associated with a two-phonon Orbach process in which a phonon is absorbed by the spin system, allowing an excited state to be accessed, with the spin system relaxing to its final state, accompanied by the emission of a phonon of a different energy.  This allows the spin system to overcome the energy barrier \cite{MorrishBook,Orbach1961}.  
An experiment such as a.c.\ susceptibility is carried out at a particular frequency $\omega$, corresponding to a particular time constant $\tau_{\rm measure}$. As $\tau(T)$ sweeps through this time constant as $T$ is lowered, the magnetization changes from dynamic (at high $T$, where $\tau(T)\ll \tau_{\rm measure}$) to frozen (at low temperature $\tau(T)\gg\tau_{\rm measure}$).  The temperature at which this crossover occurs is known as the
blocking temperature, $T_{\rm B}$, and depends entirely on the value of $\tau_{\rm measure}$ employed (thus $T_{\rm B}$ is frequency dependent).  Therefore, $T_{\rm B}$ marks the point at which relaxation becomes long for the measurements timescale.  
This principle is illustrated in the inset of figure~\ref{fig:cadiou}(b) which shows the peak of $\chi''$ moving to lower temperatures as the measurement frequency decreases.  This allows the form of $\tau(T)$ to be extracted, as shown in the main part of figure~\ref{fig:cadiou}(b).  At higher temperature (small $1/T$) $\tau$ is quite small and is measured by a.c.\ susceptibility, but it grows rapidly on cooling (following an activated dependence corresponding to thermally-assisted hopping over the energy barrier).  At low temperature (large $1/T$) a plateau in $\tau$ is observed (this is due to quantum-mechanical tunnelling through the barrier).  These very long relaxation times (large $\tau$, approaching a few hours) are measured by d.c.\ relaxation (magnetizing the sample, removing the field, and measuring the relaxation time for the magnetization to die away) rather than a.c.\ susceptibility. 

Since each molecule of a SMM is identical (and rather well magnetically isolated from its neighbour), the relaxation time of each molecule might be expected to be identical, so that the ideal Debye model should hold rather well.  In practice, there can be a range of coexisting relaxation processes, and data are better fitted by a generalized Debye model (equation~(\ref{GenDebye})) with $\alpha\neq 0$ \cite{GatteschiBook,Goura2015,Goura2016}.  Nevertheless, as the condition $\omega\tau\approx 1$ is crossed on cooling, $\alpha$ can be observed to adopt small values indicating small spreads of relaxation times and the presence of Debye-like process~\cite{Goura2016}.  The corresponding Cole-Cole plot is therefore fairly close to the idealised semicircles with only slight sinking below the $x$-axis \cite{Goura2016}.  The assumption of a symmetric spread of relaxation times implied by the generalised Debye model can be explained by appeal to the intermolecular interactions that are assumed negligible.  While each molecule should be identical and possess an identical relaxation time to every other, the inclusion of interactions could cause a smearing of relaxation times around some nominal relaxation time.  Superparamagnetic particles naturally show a (roughly) symmetric spread of nanoparticle sizes around a central mean \cite{Singh2009,Narayanasamy2018} showing a natural progression to a symmetric distribution function, as assumed in the generalised Debye model.

\begin{figure*} [tbp]
  \centering
   \fbox{\bf Copyrighted figure -- please see published article}

	\caption{Results of a.c.\ susceptibility measurements on iron oxide magnetic nanoparticles in an aqueous suspension showing the effect of coating these particles in polyethyleneimine (PEI).  Adapted from reference~\cite{Narayanasamy2018}.  The measured $\chi^{\prime\prime}$ peak was attributed to Brownian relaxation allowing the hydrodynamic size of the particles to be followed.\label{fig:ColloidSup}.}
\end{figure*}

An example of relaxation in magnetic iron oxide nanoparticles in aqueous suspension is shown in figure~\ref{fig:ColloidSup}.  Because they are suspended in fluid, the nanoparticles can physically re-orientate with a Brownian relaxation time defined by

\begin{equation}
    \tau_{\rm B}=\frac{\pi\eta D_{\rm H}^3}{2k_{\rm B}T},
\end{equation}
where $\eta$ is the dynamic viscosity of the fluid and $D_{\rm H}$ is the hydrodynamic size \cite{Narayanasamy2018}.  Coating of the particles with polyethyleneimine  changes the frequency of the $\chi^{\prime\prime}$ peak because of the change of $\tau_{\rm B}$.  The peaks associated with Arrhenius behaviour (due to moment reorientation only, see equation~(\ref{eq:arrhenius})) were found to be well separated from these Brownian relaxation peaks \cite{Narayanasamy2018}.  By way of contrast, nickel nanoparticles deposited in silica cannot physically re-orientate leading to slow relaxation controlled by magnetic moment reorientation only shown in figure~\ref{fig:SupAC} \cite{Singh2009}.

\begin{figure} [tbp]
\centering
   \fbox{\bf Copyrighted figure -- please see published article}	
	\caption{Experimental measurement results of a.c.\ susceptibility measurements of Ni nanoparticles dispersed in silica.  Adapted from reference~\cite{Singh2009}.  The average particle size $D$ is shown as derived using transmission electron microscopy\label{fig:SupAC}.}
\end{figure}

\begin{figure*} [tbp]
	\includegraphics[width = 2\columnwidth]{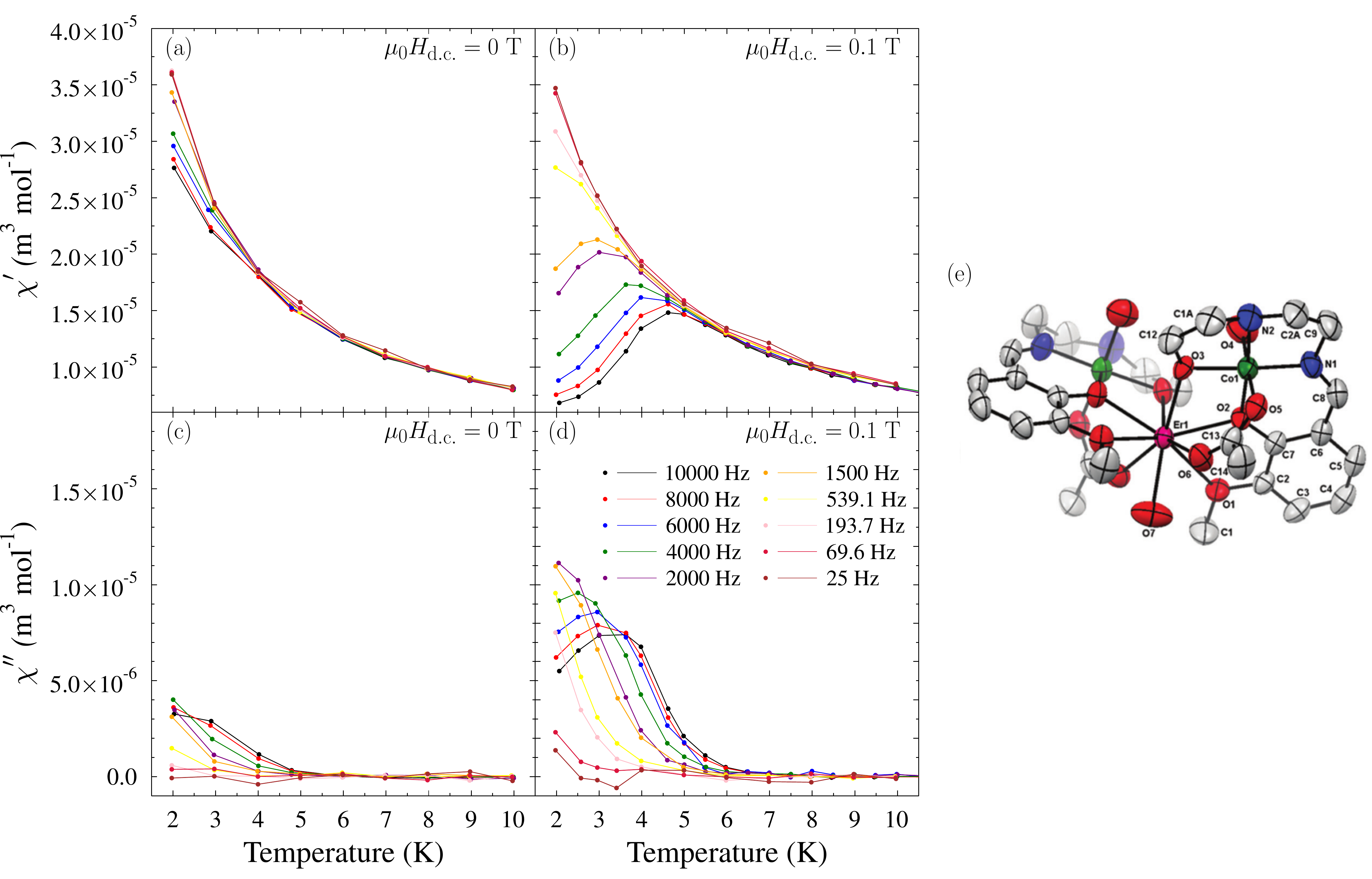}%
	
	\caption{Results of a.c.\ susceptibility measurements on [Co$_2^{\rm III}$Er(L)$_2$($\mu$-O$_2$CCH$_3$)$_2$(H$_2$O)$_3$]$\cdot$NO$_3\cdot$$x$MeOH$\cdot$$y$H$_2$O, known as Co$_2$Er, in an ac magnetic field of $\mu_0H_{\rm ac}=0.4~\rm mT$ adapted from reference~\cite{Goura2016}. 
	(In the chemical formula, LH$_3$ = 2-methoxy-6-[{2-(2-hydroxyethyl\-amino)\-ethyl\-imino}\-meth\-yl]phenol.) The (a) real and (b) imaginary parts of ac susceptibility in a d.c\ magnetic field of $\mu_0H_{\rm dc}=0~\rm T$.  The (c) real and (d) imaginary parts of ac susceptibility in a d.c\ magnetic field of $\mu_0H_{\rm dc}=0.1~\rm T$.  (e) An ORTEP diagram of Co$_2$Er from reference~\cite{Goura2016} omitting disordered parts, H atoms, anions and solvent molecules for clarity \label{Co2Er_AC}.}
\end{figure*}

\begin{figure} [tbp]
	\includegraphics[width = 1\columnwidth]{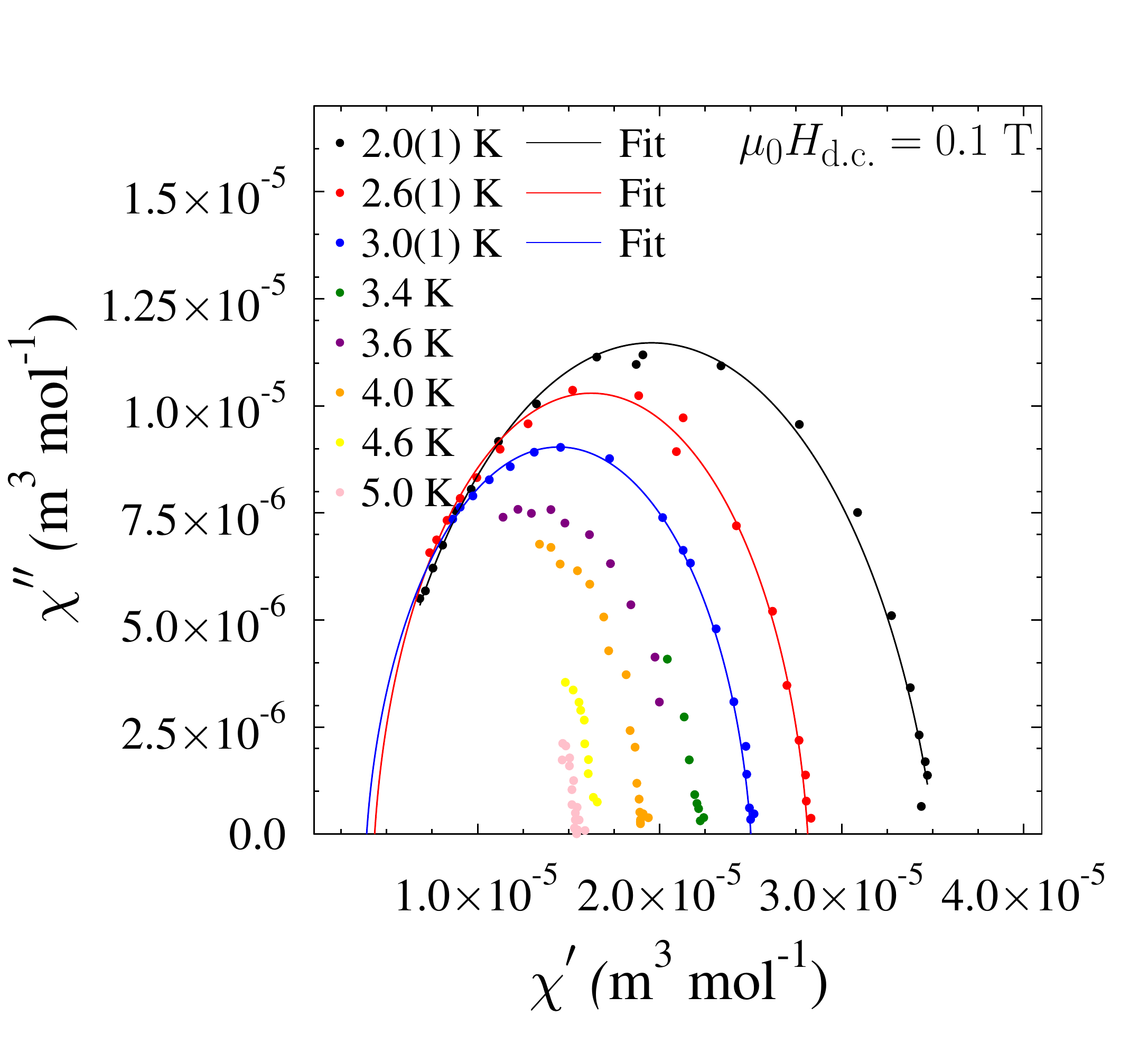}%
	
	\caption{Cole-Cole plot of Co$2$Er measurements at $\mu_0H_{\rm d.c\ c.\ }=0.1~\rm T$ showing the formation of the expected arcs for a system displaying slow magnetic relaxation from reference~\cite{Goura2016}.  Displayed fits to equation~(\ref{ColeColeTest}) are shown as lines.\label{SMMColeCole}}
\end{figure}

Figure~\ref{Co2Er_AC}(a-d) shows example a.c.\ susceptibility data for the single molecule magnet Co$_2$Er~\cite{Goura2016} and 
a diagram of the molecular structure (figure~\ref{Co2Er_AC}(e)).  Strictly this compound falls under the class of single ion magnets since only the rare earth ion possesses unpaired electrons and therefore a magnetic moment \cite{Goura2015,Goura2016}.  The Cole-Cole plot for this compound (shown in figure~\ref{SMMColeCole}) agreed well with the generalised Cole-Cole model with fits showing a low $\alpha$ of $\sim 0.2$--$0.3$ for $3$--$2~\rm K$ \cite{Goura2016}.  The fact $\alpha$ slightly increased with decreasing temperature suggested other relaxation mechanisms becoming important.  It is interesting to note that slow magnetic relaxation (evidenced by a frequency dependence in both real and imaginary susceptibilities and non-zero $\chi^{\prime\prime}$) is greatly enhanced when a d.c.\ magnetic field is applied to this compound.  In other words, it shows field-induced slow magnetic relaxation.

The role of the d.c.\ magnetic field in altering the relaxation is to open up an alternate relaxation pathway, known 
 as macroscopic quantum tunnelling or quantum tunnelling of magnetisation \cite{GatteschiBook,Gatteschi2003,Christou2000}.  This allows spins to tunnel through the energy barrier separating up and down spins in order to relax, rather than relying on a thermal process to provide the energy to leap over it \cite{GatteschiBook,Gatteschi2003,Christou2000,Blundell2004}.  Tunnelling can mask the thermal relaxation described by an Arrhenius-type equation, as evidenced by the field enhanced slow relaxation of Co$_2$Er (figures~\ref{Co2Er_AC}(c) and (d)) and other field-induced SMMs \cite{Goura2015,Goura2016,Huang2015,Miklovic2015,Biswas2016,Lin2011}.  For quantum tunnelling to occur there must be some additional term in the Hamiltonian that does not commute with $S_z$ \cite{GatteschiBook,Gatteschi2003}.  This is achieved for non-zero $E$ in equation~(\ref{SMMHamil}) \cite{GatteschiBook,Gatteschi2003}, though other higher-order terms can also contribute, all depending on the relevant symmetries of the magnetic ions in the molecule~\cite{GatteschiBook}.  Quantum tunnelling can take place when energy levels on either side of the energy barrier become degenerate and so is more likely at zero applied d.c.\ magnetic field and at higher fields corresponding to new energy levels being brought into degeneracy~\cite{Gatteschi2003,Christou2000,Blundell2004}.  These tunnelling conditions can be observed by studying the magnetic hysteresis loop which breaks up into  a series of steps at these specific fields when degeneracy is recovered and quantum tunnelling becomes more rapid~\cite{Brechin2003}.  Figure~\ref{DoubleWell}(b) and (c) illustrates the effect of increasing an applied magnetic field (assumed parallel to the $z$-axis) where the degeneracy between up and down states is progressively broken and temporarily reestablished between new levels.  
 Thus tunnelling is expected to be allowed in figures~\ref{DoubleWell}(a) and \ref{DoubleWell}(c), but forbidden in
 figure~\ref{DoubleWell}(b). 

Although the generalised Debye model can account for departures from a simple Debye model, it sweeps all the details under the carpet.  It is more profitable to try and consider the additional relaxation processes which are available in SMMs, some of which (as we have seen) can be field dependent.  One can start by considering the simplest ``direct'' process in which the transition between two levels A and B is accompanied by the absorption or emission of a phonon of energy equal to the difference in the energy of those two levels, $\delta=E_{\rm B}-E_{\rm A}$.  The direct process involves a coupling between the crystal field of the magnetic ion and the strain field produced by the phonon.  The relaxation rate of the direct process is proportional to temperature (essentially because the number of phonons available at temperature $T$ scales with $T$). The direct process (a one-phonon process) is not very efficient since the density of states of these low-energy phonons is rather low.  

Two-phonon processes allow the system to exploit more abundant higher-energy phonons.  In a two-phonon process a transition from level A to B is effected by first absorbing a phonon of energy $\Delta=E_{\rm C}-E_{\rm A}$ and exciting the system from A right up to a third (much more energetic) level C, and then emitting a phonon of energy $\Delta-\delta$ to drop down to B.  This is known as an Orbach process, which we have already described above.  The relaxation rate of an Orbach process is proportional to the Bose factor $(\exp(\Delta/k_{\rm B}T)-1)^{-1}$ which is proportional to $\exp(-\Delta/k_{\rm B}T)$ if $\Delta\gg k_{\rm B}T$ (i.e.\ recovers the Arrenhius form in equation~(\ref{eq:arrhenius}) with $E_{\rm a}=\Delta$). 

If the excited state C is virtual, it is known as a Raman process.  The detailed functional form ascribed to these different relaxation processes can depend on the nature of the magnetic ion (Kramers or non-Kramers) and the temperature and field regime being studied.   
For example, a study of both a trigonal prismatic mononuclear Co(II) complex and mononuclear hexacoordinate Cu(II) complex found slow relaxation to be due to the sum of Orbach and quantum tunnelling processes, as well as direct and Raman phonon processes, and the data were fitted to an expression given by

\begin{equation}
\tau^{-1} = AH^2T + \frac{B_1}{1+B_2H^2}+CT^n+\tau_0^{-1}\exp\left(\frac{-E_a}{k_BT}\right),\label{DifferentMechs}
\end{equation}
where $A$, $B_1$, $B_2$, $C$ and $n$ are constants and $H$ is the magnetic field~\cite{Novikov2015,Boca2017}.  These correspond to direct, quantum tunnelling of magnetisation, Raman and Orbach processes respectively.  This clearly demonstrates that a.c.\ magnetic susceptibility may have a dependence on the applied magnetic field through its effect on $\tau$.  Very often in the investigation of SMMs it is the Arrhenius type relaxation that is the relaxation of interest (in order to extract the height of the energy barrier, $E_{\rm a}$).  Thus a d.c\ field and frequency study must be performed to identify the conditions needed to minimise contributions to $\tau^{-1}$ from those field-dependent parts \cite{Goura2016,Biswas2016,Lin2011}.

These systems also provide a real example of the case of two $\tau$'s (as described by equation~(\ref{Simp2tau})).  The molecule (Cp$^*$)Er(COT) (where Cp$^*$ is C$_{5}$Me$_5^-$ and COT is C$_8$H$_8 ^{2-}$) features an Er(III) ion sandwiched between two carbon rings.  Measurements of this compound revealed two separate sets of peaks appearing in $\chi^{\prime\prime}$ at different ranges of temperatures \cite{Jiang2011}.  A.c.\ data were successfully modelled via a version of equation~(\ref{Simp2tau}) incorporating the generalised Debye model \cite{Jiang2011}.  The source of two separate $\tau$'s in this compound was suggested to be two separate conformations of (Cp$^*$)Er(COT) \cite{Jiang2011}.  Similarly, a mononuclear hexacoordinate Cu(II) complex was found to contain two $\tau$'s at low temperature which was suggested to be due to very weak intercluster interactions creating small oligomers (the discrete SMMs bonding together in ``clumps'') \cite{Boca2017}.

\subsection{Spin Glasses}\label{sect:SpinGlasses}

\begin{figure} [tbp]
	\includegraphics[width = 1\columnwidth]{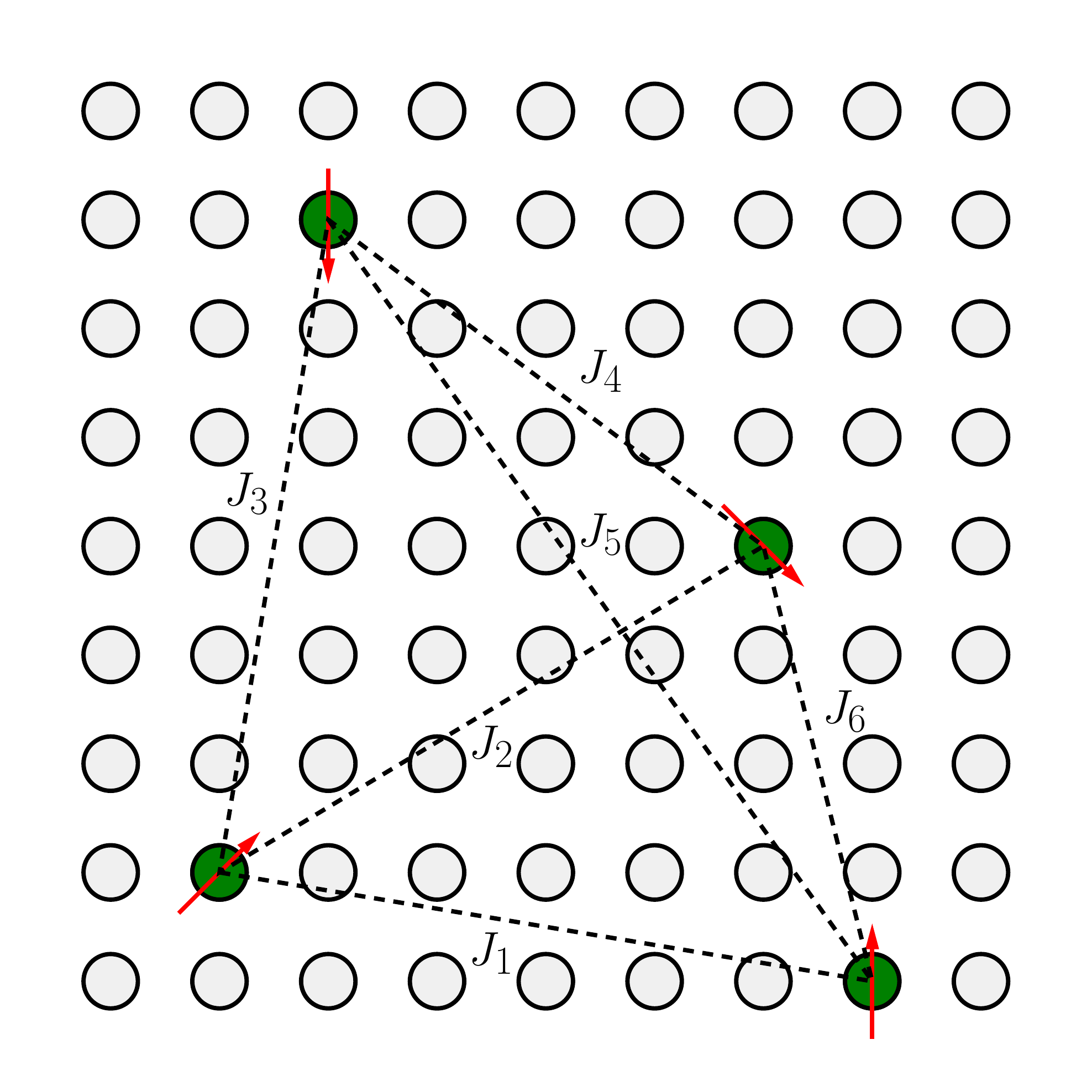}%
	
	\caption{A schematic of a typical spin glass. A low concentration of spins decorate a non-magnetic $2$-D square lattice. Due to their random locations the exchanges between each ($J_i$ for $i=1$ to $6$) are also random leading to frustration. \label{SpinGlassSketch}}
\end{figure}

A spin glass can be formed if one takes a non-magnetic lattice and populates it with a dilute random distribution of magnetic atoms, as shown in figure~\ref{SpinGlassSketch}.  An example is the CuMn system in which magnetic Mn atoms replace Cu atoms at the few per cent level~\cite{MydoshBook}.  Cu is a non-magnetic metal and so the exchange interactions between the Mn atoms are mediated through the conduction electrons by the RKKY interaction \cite{SteveBook}

\begin{equation}
    \mathcal{J}(r)\propto\frac{\cos(2k_{\rm F}r)}{r^3},
\end{equation}
where $\mathcal{J}$ is the exchange integral, $k_{\rm F}$ is the Fermi wavevector and $r$ is the distance between two Mn atoms.  $\mathcal{J}$ may be positve or negative (ferromagnetic or antiferromagnetic) depending on distance, thereby introducing frustration in the spin glasses (see figure~\ref{SpinGlassSketch}).  Spin glasses are therefore random, mixed-interacting systems which, when temperature is lowered, undergo a freezing transition from a paramagnetic state to a metastable state known as the glass or frozen state lacking in any long range magnetic order \cite{MydoshBook,Binder1986,Huang1985}.  The glassiness arises from competing interactions between individual magnetic moments and leads to a multidegenerate ground state \cite{MydoshBook}.  

This multidegenerate ground state means that the system can adopt a number of  equally favourable orientations, but upon freezing the system becomes stuck in one particular configuration.  Slow relaxations arise as individual magnetic moments begin to reorient, creating additional frustrations and further reorientation of other magnetic moments.  This process is very complex because each magnetic moment  occupies a different environment and so may be frustrated in a different way (due to the random site distribution and random exchange). As the freezing temperature, $T_{\rm f}$, is approached from higher temperatures, some moments begin to cease behaving independently and start to form growing clusters \cite{Huser1986}.  A variety of size of clusters can coexist,  leading to a wide distribution of relaxation times.

Early a.c.\ susceptibility measurements of spin glasses identified a distinctive cusp in the in-phase component, $\chi^\prime$, at $T_{\rm f}$ \cite{MydoshBook,Binder1986,Huang1985}.
Interpreting a spin glass as a collection of 
superparamagnetic clusters can allow the use of  equation~(\ref{DistribDebye}) which explicitly incorporates a distribution of relaxation times in $g(\tau)$ \cite{Lundgren1981}, often assumed to be Gaussian \cite{Heidelberg} though this does not always hold \cite{Fuoss1941}.
In this interpretation, freezing occurs when a large proportion of the superparmagnetic clusters are below their respective $T_{\rm B}$.  Sometimes data are fitted to equation~(\ref{GenDebye})  to crudely monitor the spread of relaxation times parameterized by $\alpha$ \cite{MydoshBook,Huser1986}.  This equates to assuming the distribution of relaxation times is given by equation~(\ref{eq:DistriGenDebye}) and is a sensible step in modelling should a spin glass indeed be comprised of superparamagnetic clusters (since superparamagnets with a symmetric distribution of sizes \cite{Narayanasamy2018,Singh2009} likely have a symnmetric distribution of relaxation times).   All of these approaches gloss over the details of what is actually a complex interacting system (and is emphatically not a collection of Debye relaxors).

This problem of describing a spin glass as a collection of (non-interacting) superparamagnetic clusters can be examined by considering the nonlinear, third harmonic of the susceptibility.  Both the alloy Cu$_{97}$Co$_3$ and spin glass Au$_{96}$Fe$_4$ show similar linear susceptibility cusps but differ in their real third harmonic \cite{Bitoh1996}.  It was found that this harmonic (expected as a negative divergence above and below the freezing temperature \cite{Fujiki1981,Suzuki1977}) could be modelled as a distribution of superparamagnetic particles for Cu$_{97}$Co$_3$ but not for Au$_{96}$Fe$_4$ suggesting that this could be used to differentiate the two behaviours \cite{Bitoh1996}.  This study did not vary frequency or a.c.\ amplitude so it is unclear what frequency dependence (if any) exists in these results.

\begin{figure} [tbp]
	 \fbox{\bf Copyrighted figure -- please see published article}
	\caption{Example a.c.\ susceptibility of spin glass (Eu$_{0.2}$Sr$_{0.8}$)S at $\mu_0H_{\rm d.\ c.}=0~\rm T$ and $\mu_0H_{\rm a.\ c.}=0.01~\rm mT$ from reference~\cite{Huser1983}.  Filled shapes correspond to $\chi^\prime$ while empty refer to $\chi^{\prime\prime}$ with $\circ = 10.9~\rm Hz$, $\square = 261~\rm Hz$ and $\triangle = 1969~\rm Hz$.\label{Huser_SpinGlass_ACExamp}}
\end{figure}

\begin{figure} [tbp]
	 \fbox{\bf Copyrighted figure -- please see published article}
	\caption{Cole-Cole plots of (Eu$_{0.2}$Sr$_{0.8}$)S at $\mu_0H_{\rm d.\ c.}=0~\rm T$ and $\mu_0H_{\rm a.\ c.}=0.01~\rm mT$ from reference~\cite{Huser1983}.  Numbers in the plot correspond to a.c.\ drive frequencies.  Lines are a result of fits to the data assuming data to be symmetric. \label{Huser_SpinGlass_ColeExamp}}
\end{figure}

As an example of a typical spin glass, susceptibility data~\cite{Huser1983} for the spin glass (Eu$_{0.2}$Sr$_{0.8}$)S are shown in figure~\ref{Huser_SpinGlass_ACExamp}.  The Cole-Cole plot of this data (figure~\ref{Huser_SpinGlass_ColeExamp}) shows strongly depressed semi-circles suggesting a broad distribution of relaxation times. In fact, the broader the distribution of relaxation times the broader and more rounded the expected $\chi^{\prime\prime}$ peak.  This peak occurs when the average relaxation time $\tau_{\rm avg}$ matches $\omega^{-1}$. 

Since $\tau_{\rm avg}$ is temperature dependent, the form of  $\tau_{\rm avg}(T)$ can be used to try to identify the type of relaxation. Although an Arrhenius expression can appear to be successful in modelling $\tau(T)$ (and would make physical sense if a spin glass were composed of identical superparamagnetic clusters), it invariably yields unphysical values of the parameters $E_{\rm a}$ and $\tau_0$ \cite{MydoshBook,Huang1985,Huser1983}.  The interacting nature of spin glasses is better accounted for using the Vogel-Fulcher law 
\begin{equation}
\tau = \tau_0 \exp\left[ \frac{E_{\rm a}}{k_{\rm B}(T-T_0)}\right]
\end{equation}
which is better suited to glass-like systems \cite{Huang1985,Huser1983,Tholence1980}.  Here, $E_{\rm a}$ and $\tau_0$ remain the activation energy and characteristic relaxation time while $T_0$ is a new parameter that accounts for the interactions occurring between moments in a spin glass \cite{Balanda2013,Shtrikman1981}.  It should be noted that $\tau$ may be modelled by further equations~\cite{Balanda2013,Souletie1985}.  
The frequency dependence of the extracted freezing temperature (the temperature at which $\chi'$ takes its maximum) is given by $f={\rm d}\ln T_{\rm f}/{\rm d}\ln\omega$.  (Of course one can equivalently write $f={\rm d}\log_{10}T/{\rm d}\log_{10}\omega$.)  In experiments $T_{\rm f}$ changes by a small amount as $\omega$ changes over several orders of magnitude, and so $f$ can be estimated using
\begin{equation}
f=\frac{\Delta T_{\rm f}}{T_{\rm f}\Delta [\ln(\omega)]}.\label{DeltaTf}
\end{equation}
If $f$ is a constant then  $T_{\rm f}\propto\omega^f$.
Spin glasses typically give a value of $f$ between $0.001$ and $0.08$ (much larger values are found for single molecule magnets \cite{GatteschiBook,MydoshBook}) and thus $T_{\rm f}$ has a very weak frequency dependence. 

While the discussion of this section has been concerned with the spin glass state emerging out of the paramagnetic state as temperature is lowered, spin glass-like states (with associated slow magnetic relaxation) have been studied emerging in other scenarios.  The reentrant spin glass is one such example where a spin glass-like freezing occurs below a ferromagnetic transition \cite{Binder1986,Verbeek1978,Yeshurun1980} though it has been suggested this is not a true spin glass state \cite{Abiko2005}.  The d.c.\ magnetisation increases as temperature is lowered due to the appearance of the ferromagnetic state, but decreases at even lower temperatures due to freezing \cite{Abiko2005,Aslibeiki2009}.  Figure~\ref{fig:ReentrantSGAC} shows the $\chi^{\prime\prime}$ peaks associated with the reentrant spin glass La$_{0.8}$Sr$_{0.2}$Mn$_{0.925}$Ti$_{0.075}$O$_3$.  Another example is the low temperature spin-glass like state that occurs due to the presence of iron in the lithium hydroxide layers of Li$_{1-x}$Fe$_x$(OH)Fe$_{1-y}$Se \cite{Woodruff2016,Topping2017}.  This compound displays spin glass-like slow magnetic relaxation that persists in the presence of the superconductivity in the iron selenide layers \cite{Topping2017}.

\begin{figure} [tbp]
\centering
	 \fbox{\bf Copyrighted figure -- please see published article}
	\caption{Imaginary part of the a,c,\ susceptibility of La$_{0.8}$Sr$_{0.2}$Mn$_{0.925}$Ti$_{0.075}$O$_3$ demonstrating a reentrant spin glass.  The peaks at $\sim 170~\rm K$ correspond a ferromagnetic transition while the lower set are associated with the slow relaxation of the frozen state.  The inset highlights these peaks.  Adapted from reference~\cite{Aslibeiki2009}. \label{fig:ReentrantSGAC}}
\end{figure}

\subsection{Spin Ice}\label{sect:SpinIce}

As with spin glasses, spin ices show slow magnetic dynamics.  However, while intrinsic randomness is important in explaining these dynamics in spin glasses, a different explanation is needed for spin ices which have a completely ordered chemical composition.  Spin ice behaviour was first identified in certain  rare-earth pyrochlores \cite{Harris1997}.  Pyrochlores have formula A$_2$B$_2$O$_7$ and contain a rare earth A which resides on the vertices of a network of corner-sharing tetrahedra \cite{Harris1997, Bramwell2001}.  When A = Dy or Ho and B = Ti, the crystal field associated with the A cations exhibits strong Ising anisotropy that causes each A spins to lie along the line joining the centres of the two neighbouring tetrahedra which share the corner occupied by the A cation (these lines are along the $\langle 111\rangle$ directions).  Each spin can lie parallel or antiparallel to this line.  The dipolar interaction and weak antiferromagnetic superexchange results in an effective ferromagnetic interaction.  This, in combination with the single-site anisotropy, results in a spin arrangement that is subject to a constraint on each tetrahedron such that its four spins, one on each corner, should satisfy $\sum_i \vec{S}_i = 0$.  Thus two of the spins point into the centre of each tetrahedron while the other two point out, akin to the proton rules of water ice \cite{Anderson1956, PaulingBook}.
There are a large number of ways of satisfying this constraint and so this results in a degenerate ground-state configuration (lacking long-range order down to low temperatures)~\cite{Harris1997, Ryzhkin2005}. 

A.c.\ magnetic susceptibility measurements on spin ice show an Arrhenius-like behaviour at high temperature (due to single ion processes and mixing with excited states) but on cooling the dynamics begin to freeze out \cite{Snyder2004}. 
This behaviour can look superficially like that of spin glasses, with  a low temperature peak in d.c.\ susceptibility accompanied by a divergence between ZFC and FC sweeps.  This is expected as in both cases a slowing down of spin dynamics is occurring.  

The low-temperature excitations of spin ice have attracted considerable attention.  A single spin-flip breaks the 2-in, 2-out constraint if the spin ice, creating  a ``three in/one out'' and ``three out/one in'' pair of tetrahedra.  This state can be considered as a monopole-antimonopole pair \cite{Ryzhkin2005,Castelnovo2008} and further spin flips allow the monopole and antimonopole to separate, travelling through the lattice (the initial spin flip is said to have been ``fractionalised'').   Slow magnetic dynamics in the low-temperature state of a spin ice system can therefore be described by considering monopole motion.  In fact, an early description   of spin ice magnetic relaxation by Ryzhkin~\cite{Ryzhkin2005} was formulated by exploiting the analogy with dielectric relaxation in water ice.  In this approach, the magnetic monopole current density $\boldsymbol{J}=\partial\boldsymbol{M}/\partial t$ can be written as
\begin{equation}
    \boldsymbol{J}=\kappa(\boldsymbol{H}-\chi_{\rm T}^{-1}\boldsymbol{M}),
    \label{eq:ryzhkin}
\end{equation}
where $\kappa$ is the monopole conductivity \cite{Ryzhkin2005,Bramwell2012}. In equilibrium, $\boldsymbol{M}=\chi_{\rm T}\boldsymbol{H}$ and $\boldsymbol{J}=0$.  Out of equilibrium, the monopole current contains two terms, the familiar drift term and the more unusual reaction field (originating from configurational entropy of the monopole vacuum i.e.\ the statistical mechanics of the spin configurations subject to topological constraints), and these two terms will not cancel.  To understand the time dependence, it is helpful to Fourier transform equation~(\ref{eq:ryzhkin})
which results in 
\begin{equation}
    \tilde{\boldsymbol{J}}=-i\omega\tilde{\boldsymbol{M}} =
    \kappa(\tilde{\boldsymbol{H}}-\chi_{\rm T}^{-1}\tilde{\boldsymbol{M}})
\end{equation}
and hence the magnetic susceptibility $\chi(\omega)=\tilde{M}/\tilde{H}$ 
can be written as
\begin{equation}
\chi(\omega)= \frac{\chi_{\rm T}}{1-i\omega \tau},
\label{eq:spinice-debye}
\end{equation}
where $\tau=\chi_{\rm T}/\kappa$; this is clearly Debye-like.
We note that this approach can be extended to spatial deviations, resulting in a diffusion term being added to equation~(\ref{eq:ryzhkin}) as follows: 
\begin{equation}
    \boldsymbol{J}(\boldsymbol{r}) =\kappa[\boldsymbol{H}(\boldsymbol{r})-\chi_{\rm T}^{-1}\boldsymbol{M}(\boldsymbol{r})]+D\nabla^2 \boldsymbol{M}(\boldsymbol{r}).
    \label{eq:ryzhkin2}
\end{equation}
Another approach \cite{Armitage} is to include a phenomenological inertial term into equation~(\ref{eq:ryzhkin}) to model a relaxation of the monopole current and this leads to 
\begin{equation}
\chi(\omega)= \frac{\chi_{\rm T}}{1-i\omega \tau-\omega^2\tau/\gamma},
\label{eq:spinice-debyeAccel}
\end{equation}
where $\gamma$ is a monopole current relaxation rate. This equation (which is equivalent to equation~(\ref{Accel})) has been used to model data obtained in the quantum spin ice Yb$_2$Ti$_2$O$_7$ \cite{Pan2015}.

\begin{figure} [tbp]
\centering
	 \fbox{\bf Copyrighted figure -- please see published article}
	\caption{Results of low temperature a.c.\ susceptibility measurements on spin ice Dy$_2$Ti$_2$O$_7$ from reference~\cite{Matsuhira2001} at  $\mu_0H_{\rm d.\ c.}=0~\rm T$ and $\mu_0H_{\rm a.\ c.}=0.5~\rm mT$.\label{Matsuhira_SpinIce_ACExamp}}
\end{figure}

\begin{figure} [tbp]
	\includegraphics[width = 1\columnwidth]{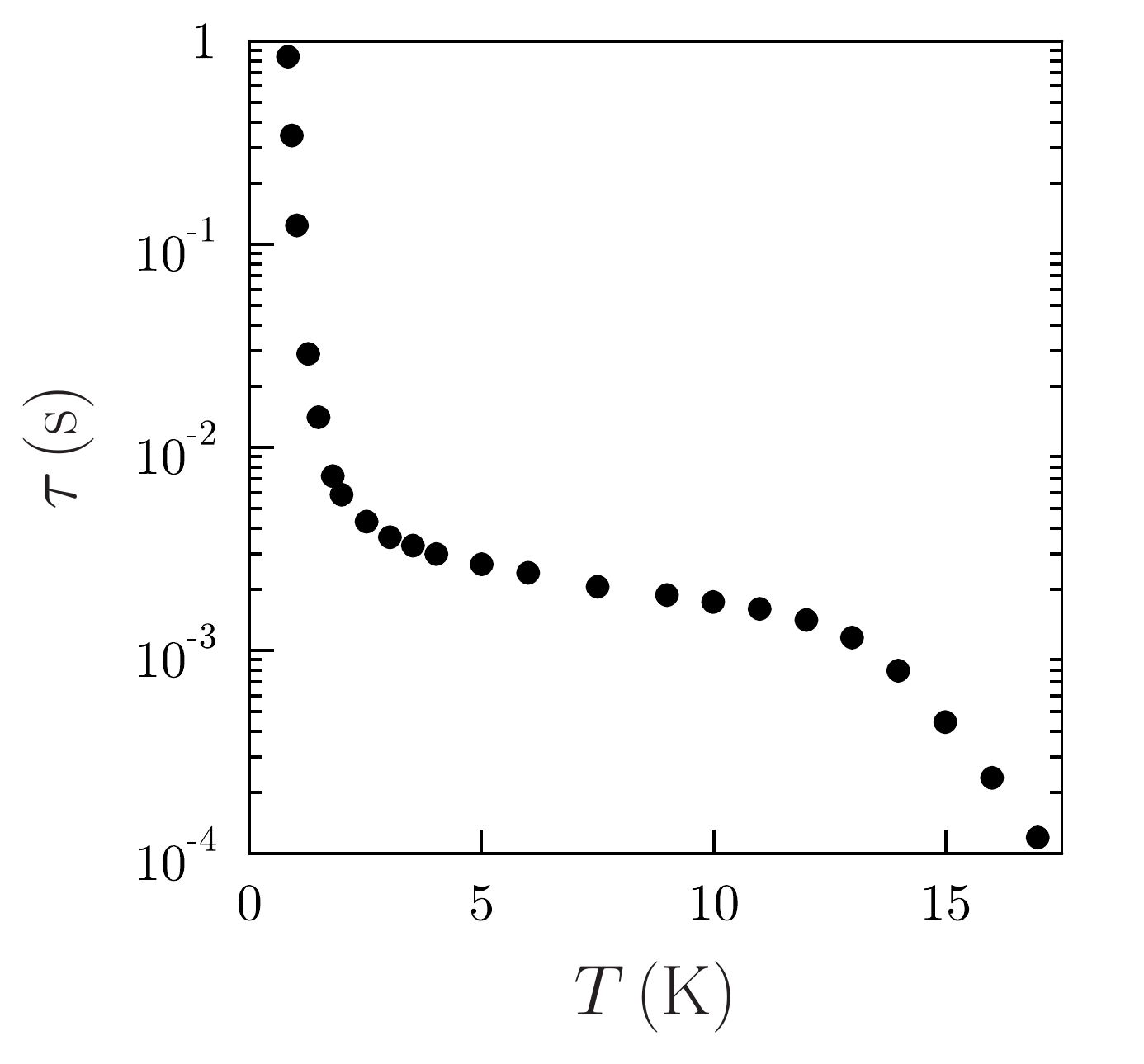}%
	\caption{The extracted $\tau(T)$ for Dy$_2$Ti$_2$O$_7$ from the data of \cite{Snyder2004} showing activated behaviour at $T>15$\,K corresponding to crystal field excitations, below which there is a quasi-plateau region and these excitations freeze out, and then a low-temperature increase in $\tau$ as the sample enters the spin-ice regime. \label{fig:ACspinice}}
\end{figure}

We now consider the specific example of Dy$_2$Ti$_2$O$_7$, which is one of the most highly studied examples of a classical spin ice. The low temperature ($\sim 1$~K) peak in $\chi'$ shown in figure~\ref{Matsuhira_SpinIce_ACExamp} matches the freezing transition appearing in the d.c.\ susceptibility.  However, a.c.\ susceptibility reveals a peak in $\chi''$ at higher temperature (up to $\sim 20$~K) measured at high $\omega$ \cite{Matsuhira2001,Snyder2004} (not shown here).  
The arcs in the Cole-Cole plots in figure~\ref{Matsuhira_SpinIce_Cole} allow these data to be modelled and the average relaxation time $\tau(T)$ can be extracted, as shown in figure~\ref{fig:ACspinice}.  The clearly asymmetric arcs were modelled with the Cole-Davidson model featuring a maximum cut-off relaxation time and a long low-$\tau$ tail \cite{Matsuhira2001}.  The origin of this behaviour was unclear beyond being related to the spin ice state \cite{Matsuhira2001}.  
The relaxation time
increases on cooling in the thermally activated high-temperature regime (associated with excitations to higher crystal field levels), entering a plateau region below around 12 K (associated with quantum tunnelling processes through the crystal field barrier), before experiencing a sharp upturn below around 2 K (associated with monopole dynamics), explainable only by including the Coulomb interaction between charges \cite{JaubertHoldsworth}.

\begin{figure} [tbp]
\centering
	 \fbox{\bf Copyrighted figure -- please see published article}
	\caption{Cole-Cole plot of frequency dependent feature of Results of Dy$_2$Ti$_2$O$_7$ from reference~\cite{Matsuhira2001}.  
	$\mu_0H_{\rm a.\ c.}=0.5~\rm mT$ and $\chi^{\prime}_{\rm int}$ and $\chi^{\prime\prime}_{\rm int}$ indicate $\chi^{\prime}$ and $\chi^{\prime\prime}$ corrected for the demagnetizing factor. Panel (b) is a blow-up of panel (a), while panel (c) shows high temperature fits to the Cole-Davidson model (equation~(\ref{Cole_Davidson})).\label{Matsuhira_SpinIce_Cole}}
\end{figure}

\subsection{Long Range Magnetic Order}

Our discussion so far has focused on systems lacking long range order.  However, slow magnetic relaxation can also be observed in systems with long range order if they contain some component which responds rather sluggishly. The pertinent question is: ``What is the sluggish entity?''  Consider first a ferromagnet in which each spin is coupled to a neighbour by an exchange constant $J$.  The dynamics of the spins occurs on a frequency scale given by $J/\hbar$ which is always far in excess of anything detectable by a.c.\ susceptibility (and hence spin waves are studied using neutron scattering, see figure~\ref{FreqRange}). However, the simple, fully-aligned ferromagnetic state is not usually obtained during a ZFC susceptibility measurement because the sample breaks up into domains.  

Domain walls in ferromagnets (and also ferrimagnets), which are structures much larger than a single spin, are  more easily affected by a slowly oscillating magnetic field (which can rock the domain wall back and forth) and  will have some characteristic timescale yielding a non-zero $\chi^{\prime\prime}$ detectable in a.c.\ measurements \cite{Balanda2013}.   Various types of domain wall movement can be measured, including dynamic wall pinning and depinning, as well as domain structure reconstruction (which has been detected in a study of Sm$_2$Fe$_{17}$ \cite{Chen1996}) and irreversible magnetisation rotation \cite{Balanda2013}.  
The Curie temperature $T_{\rm C}$ can be used as a rough estimate of the size of the magnetic exchange, and so for materials with high $T_{\rm C}$ (and hence large $J$) one may need very high frequencies (at the top of the a.c.\ suscpetibility range) in order to observe the frequency-dependent relaxation \cite{Balanda2013}.

In antiferromagnets, one cannot drive a domain wall using an oscillating magnetic field and so there is not a mechanism for the moments to absorb energy.  Thus an antiferromagnetic transition may be detected via a.c.\ susceptibility in $\chi^\prime$ but will be absent in $\chi^{\prime\prime}$ showing no dissipation \cite{Balanda2013}.  This does offer a method for differentiating between ferro/ferrimagnetism and antiferromagnetism using a.c.\ susceptibility.

However, the measurement of higher harmonic a.c.\ susceptibility in antiferromagnets has proven useful in revealing previously hidden magnetic phases.  The antiferromagnet RuSr$_2$GdCu$_2$O$_8$ gave rise to additional features in the real part of the third harmonic slightly above $T_{\rm N}$ \cite{Cimberle2006} (positive above and below $T_{\rm N}$ suggesting antiferromagnetism \cite{Fujiki1981}).  This was identified as evidence of ferromagnetic nanonclusters within magnetic grains, confirmed by the d.c.\ measurement of the decay of the magnetisation below the superparamagnetic blocking temperature \cite{Cimberle2006}. Higher harmonics are also important in some incommensurate materials which can exhibit a nonlinear response to an a.c.\ magnetic field \cite{Clements2018}.  Such nonlinear responses can be described by a nonlinear sping model \cite{Mito2015,Tsurata2018}.

\begin{figure} [tbp]
	\includegraphics[width = \columnwidth]{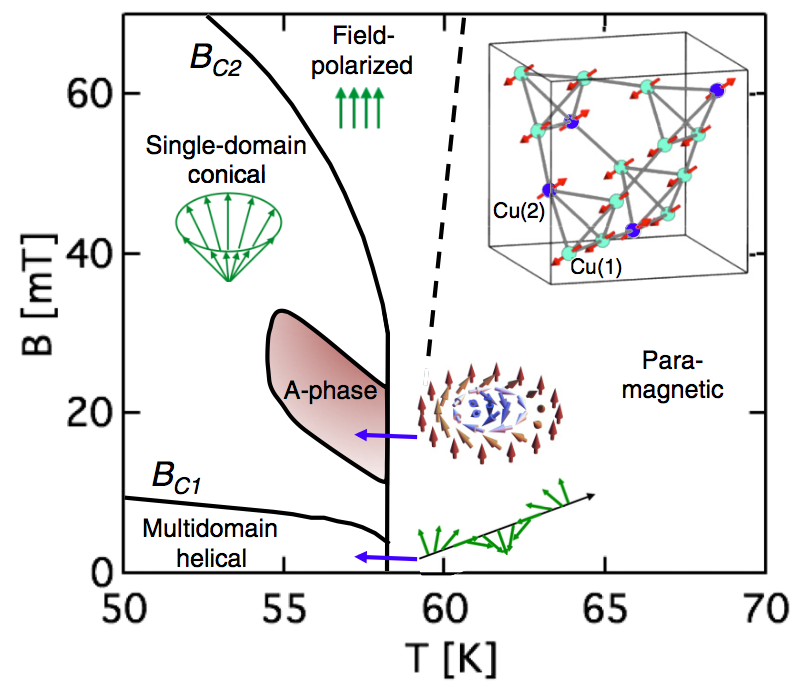}%
	\caption{The phase diagram of Cu$_2$OSeO$_3$ illustrating the
          spin arrangements of the various phases. The skyrmion is the
          A-phase. The inset shows the crystal structure.
          (After \cite{qian2016} and appears in arXiv:1607.08177.)
	\label{fig:skyrmion-pd}}
\end{figure}

\begin{figure*} [tbp]
	\includegraphics[width = 2\columnwidth]{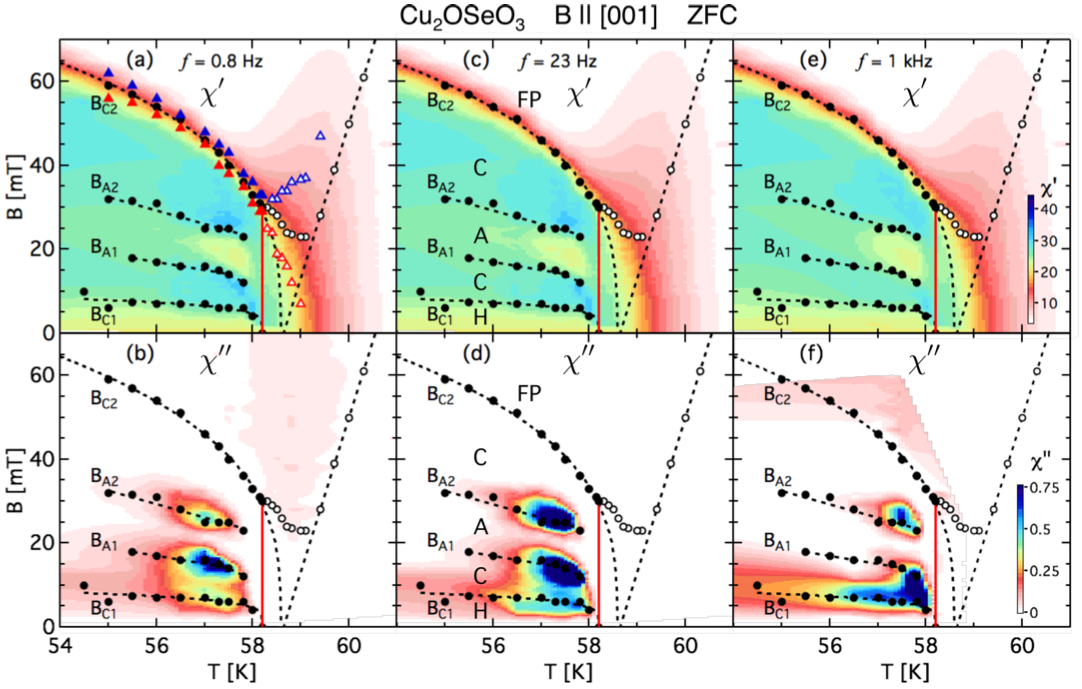}%
	\caption{Contour plots of zero-field cooled $\chi'$ and $\chi''$ measured in Cu$_2$OSeO$_3$ at various frequencies measured as a function of temperature and magnetic field. Above $T_{\rm c}$ the extrema of the first and second derivatives of $\chi'$ are illustrated with open symbols to distinguish from $B_{\rm C2}$ below $T_{\rm c}$. In panels (c) and (d) the phases are: H=helical, C=conical, A=A phase, and FP=field polarized phase.  (After \cite{qian2016} and appears in arXiv:1607.08177.)
	\label{fig:skyrmion}}
\end{figure*}

We now turn to a more complicated type of magnetic order.  Magnetic skyrmions are spin textures which are topological  \cite{rossler2006}.  These whirling magnetic structures can occur in two-dimensional crystals of chiral magnets and are of interest for potential magnetic storage applications. 
An example is the compound Cu$_2$OSeO$_3$ which exhibits a skyrmion phase (known as the A-phase) that is stabilised in a rather small region of the magnetic field--temperature ($B$--$T$) phase diagram (see figure~\ref{fig:skyrmion-pd}).
In this material there is a balance between the ferromagnetic exchange and the Dzyaloshinskii-Moriya interaction which can lead to long-period helical order, with the helices fixed along a particular crystallographic direction below $T_{\rm c}\approx 58$\,K.   When $B>B_{\rm c1}$ the helices can be oriented along $B$ leading to a conical phase, with higher fields stabilising the A-phase (and much higher fields leading to the field polarized phase)

In a.c.\ susceptibility measurements $\chi^{\prime}$ is found to follow the static susceptibility, but near phase boundaries $\chi^{\prime\prime}$ it becomes strongly frequency-dependent. This effect has been studied in various chiral magnets, including MnSi \cite{bauer2012}, GaV$_4$S$_8$ \cite{butykai1996},
Fe$_{1-x}$Co$_x$Si \cite{bannerberg2016}  
and Cu$_2$OSeO$_3$ \cite{levatic2014,qian2016}, and can be used to rather efficiently map out the phase boundaries and thus determine the phase diagram.  An example is given in figure~\ref{fig:skyrmion} \cite{qian2016} which displays a.c.\ susceptibility data on contour plots using the same $B$--$T$ phase diagram as figure~\ref{fig:skyrmion-pd}.
This figure illustrates that a lot of information is obtainable by measuring a.c.\ susceptibility carefully over a restricted range of $B$ and $T$.  It is noticeable that there is not much frequency dependence in $\chi^{\prime}$ (panels (a), (c) and (e) of figure~\ref{fig:skyrmion} are very similar), while $\chi^{\prime\prime}$ (panels (b), (d) and (f)) exhibits a much stronger frequency dependence.  Moreover, $\chi^{\prime\prime}$ shows up most strongly near (first-order) phase boundaries where dissipation occurs (see also \cite{levatic2014}).  The response in these different phases is not very well understood at a quantitative level, but such experiments demonstrate that a.c.\ susceptibility is a very powerful probe of the complex phase diagrams found in chiral magnets.  It seems plausible that a skyrmion can be imagined as a type of domain wall wrapped around in a loop and thus the slow dynamics in these phases might be relatable to those found in ferromagnets.  The strong frequency dependence in $\chi^{\prime\prime}$, which can be studied using Cole-Cole plots \cite{qian2016,butykai1996,bannerberg2016}, also points to glassy-type behaviour originating in collective dynamics.

\subsection{Superconductors}
\label{sec:superconductor}
Although superconductors do not carry distinct magnetic moments, they can nevertheless show a strong magnetic response. In the Meissner state, currents running around the edge of a sample serve to screen the interior from magnetic flux, resulting in $\chi_{\rm dc} = -1$ (in SI units) \cite{AnnettBook}.  Because the superconducting currents are  dissipationless (zero resistance) one would expect that $\chi^{\prime\prime}=0$, except perhaps very close to the critical temperature, $T_{\rm c}$, where the small oscillating magnetic field may drive the sample between superconducting and normal states.  Type~I superconductors exhibit only the Meissner state for $H<H_{\rm c}(T)$ and $T<T_{\rm c}$, otherwise they are in the normal state.  Type II superconductors exhibit two critical fields and for $H_{\rm c1} < H < H_{\rm c2}$ a mixed state exists \cite{AnnettBook}.  In this mixed state, magnetic flux may partially penetrate the superconductor in the form of flux lines which arrange themselves in the regular Abrikosov flux lattice \cite{AnnettBook}.  Thus, we would expect these flux lines to be the magnetic entities that may relax slowly.  Type II superconductors are the most widely studied and most heavily used in applications; thus the remaining discussion will focus on them.

If a current density $\boldsymbol{J}$ flows in a superconductor in a magnetic field $\boldsymbol{B}$ then a Lorentz force of density $\boldsymbol{J}\times\boldsymbol{B}$ acts on the flux lines.  If these line vortices move at velocity $\boldsymbol{v}_{\rm L}$, then a transverse electric field $\boldsymbol{E}=\boldsymbol{B}\times\boldsymbol{v}_{\rm L}$ is produced by Faraday's law.  Moving flux lines dissipate energy (due to effects of the changing magnetic field on normal state carriers as the vortex cores trundle past, as well as pair-breaking and pair-repairing at the front and back of each moving core) and this can be modelled as a viscous drag force density $\eta\boldsymbol{v}_{\rm L}$; when this balances $\boldsymbol{J}\times\boldsymbol{B}$, the vortex velocity $\boldsymbol{v}_{\rm L}$ becomes constant.  A non-zero electric field in a superconductor implies dissipation because the work done (per unit volume, per unit time) is $\boldsymbol{J}\times\boldsymbol{B}\cdot\boldsymbol{v}_{\rm L}=\boldsymbol{J}\cdot\boldsymbol{E}$ and so the electric field due to the movement of flux lines can be imagined as resulting from an effective resistivity, termed the flux-flow resistivity.  The movement of flux lines can be opposed by a pinning force which is produced by defects and inhomogeneities in the superconductor.  Near these pinning centres, the superconducting order parameter is strongly depressed, so they are ideal sites at which to locate a flux line, in the core of which the superconducting order parameter is depressed anyway (this type of pinning is known as core pinning).  Pinning centres therefore exert an attractive interaction on the flux line.  If the Lorentz force exceeds the pinning force, the flux lines can flow (this is known as flux flow).  Otherwise, flux flow is forbidden, though there is a thermally activated process (known as flux creep) that can still occur.  Both flux flow and flux creep lead to dissipation (because of the electric field associated with their movement) and therefore can contribute to $\chi''$.
Because of these processes, the
 a.c.\ response of type II superconductors is not solely dependent on frequency (an assumption we have made for previous materials) but also depends on the amplitude of the a.c.\ field applied \cite{Nikolo1994, Gomory1997, Nikolo2016, Nikolo2015, VanderBeek1993,Qin1996, Ishida1990, HandbookSupMat}, as well as on the strength of any d.c.\ field which might also be present.
 
Let us consider a number of possible regimes.
First, with a low amplitude of the a.c.\ field but with a larger d.c.\ field also applied one can often find that the flux lines are weakly pinned and the response due to their viscous motion is linear \cite{Gomory1997}.  The linear flux flow resistivity, $\rho_{\rm FF}$, is then related to the normal state resistivity, $\rho_{\rm n}$, by the empirical equation,

\begin{equation}
   \rho_{\rm FF}=\rho_{\rm n}\frac{B}{B_{\rm C2}}\simeq \rho_{\rm n}\frac{B_{\rm d.c.}}{B_{\rm C2}},
\end{equation}
where we have assumed that $B_{\rm d.c.}\gg B_{\rm a.c.}$ such that the a.c.\ field is essentially a negligible contribution to the overall field \cite{Gomory1997,HandbookSupMat}.  This resistivity feeds into the expression for the penetration depth of the a.c.\ field $\delta=\sqrt{2\rho_{{\rm FF}} /\mu_0 \omega}$ and can be related to to an Arrhenius type equation for thermally activated flux flow \cite{Gomory1997,VanderBeek1993} which will be discussed below.  To see how this works, consider a flat slab of superconductor (which we will treat as an infinite slab) of width $d$ and with flux flow resistance $\rho$ and apply an oscillating field $B_0{\rm e}^{i\omega t}$.  This oscillating field penetrates in from both sides, yielding a field profile
\begin{figure} [tbp]
\centering
	\includegraphics[width = 1\columnwidth]{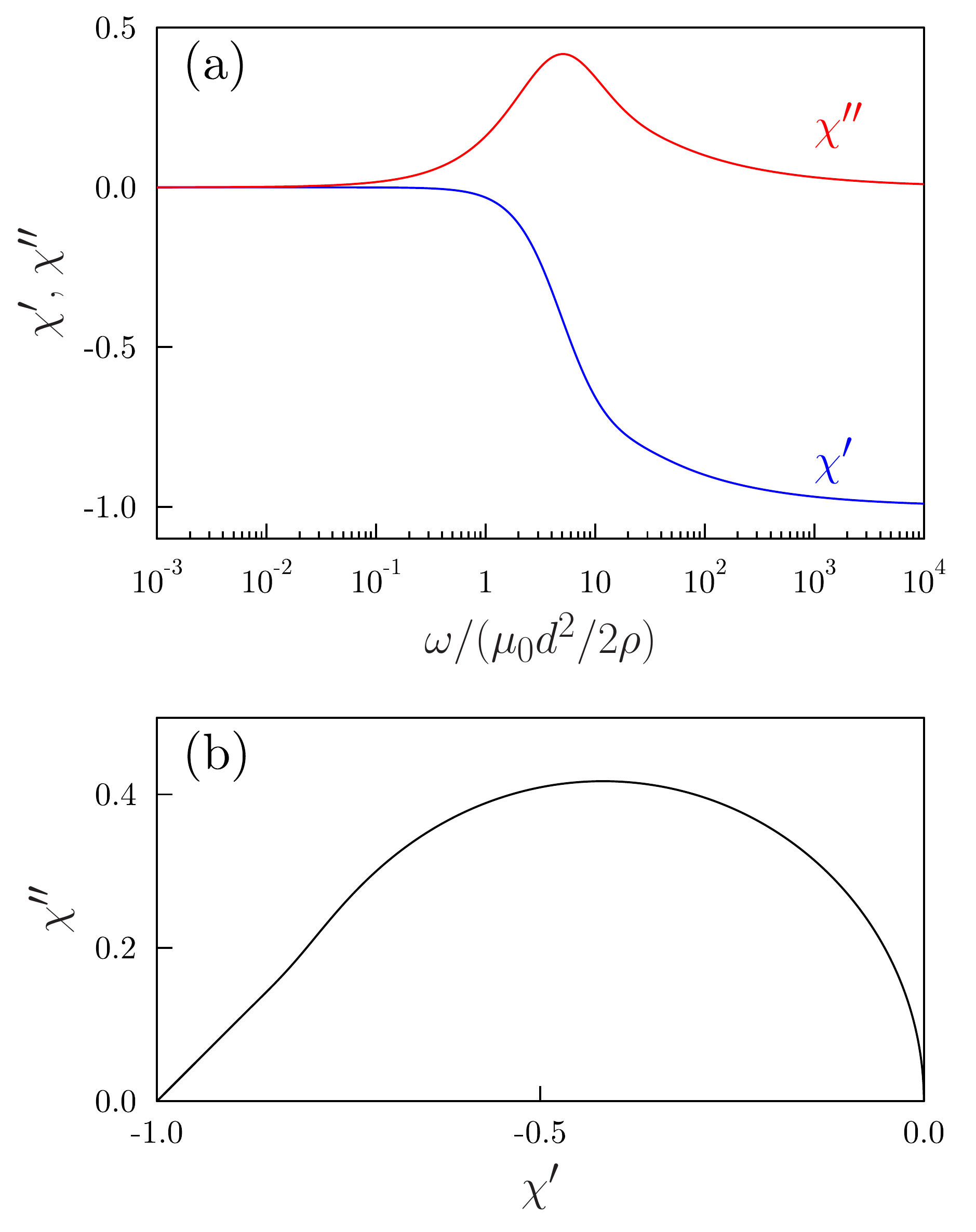}%
		\caption{(a) $\chi'$ and $\chi''$ and  (b) the Cole-Cole plot for the linear diffusion model calculated for a slab of width $d$, following equations~(\ref{eq:linearchi1}) and (\ref{eq:linearchi2}), as derived in~\cite{geshkenbein1991} .\label{fig:lineardiffusion}}
\end{figure}
\begin{equation}
    B(x,t)=B_0{\rm e}^{i\omega t}b(x),
\end{equation}
where the function $b(x)$ is given by
\begin{equation}
    b(x) = {{\rm e}^{ikx}+{\rm e}^{ik(d-x)} \over 1+{\rm e}^{ikd}}
\end{equation}
and $k=(1+i)/\delta$.  The magnetization $M(x,t)=\mu_0^{-1}B(x,t)-H(x,t)$ and hence the measured susceptibility $\chi=\langle M/H \rangle_x$ is given by averaging $M(x,t)$ over the sample, i.e.\ by evaluating
\begin{equation}
    \chi = \left( {1 \over d} \int b(x)\,{\rm d}x \right) - 1.
\end{equation}
This leads to
\begin{eqnarray}
  \chi' & = & { \sinh u + \sin u  \over u(\cosh u + cos u) } -1 \label{eq:linearchi1} \\
 \chi'' & = & { \sinh u - \sin u  \over u(\cosh u + cos u) },
  \label{eq:linearchi2}
\end{eqnarray}
where $u=d/\delta$ and $\delta=(2\rho/\mu_0\omega)^{1/2}$ \cite{geshkenbein1991}.  This is plotted in figure~\ref{fig:lineardiffusion} and can be understood as follows.  For low frequency, $\delta$ is very large and the field penetrates completely, resulting in $\chi''\to 0$.  At high frequency, $\delta$ is very short and the field only penetrates into the surface; there is hence almost complete screening and $\chi'\to -1$ and $\chi''\to 0$.  The largest absorption of energy occurs at intermediate frequencies (the maximum in $\chi''$ occurs for $\omega/(\mu_0d^2/2\rho)\approx 5.1$).  Note that for this model, the behaviour is controlled not only by adjusting $\omega$, but also $\rho$, which for flux-flow resistance is controlled by the d.c.\ field, giving another variable for the experimentalist to play with.

\begin{figure} [tbp]
\centering
	\includegraphics[width = 1\columnwidth]{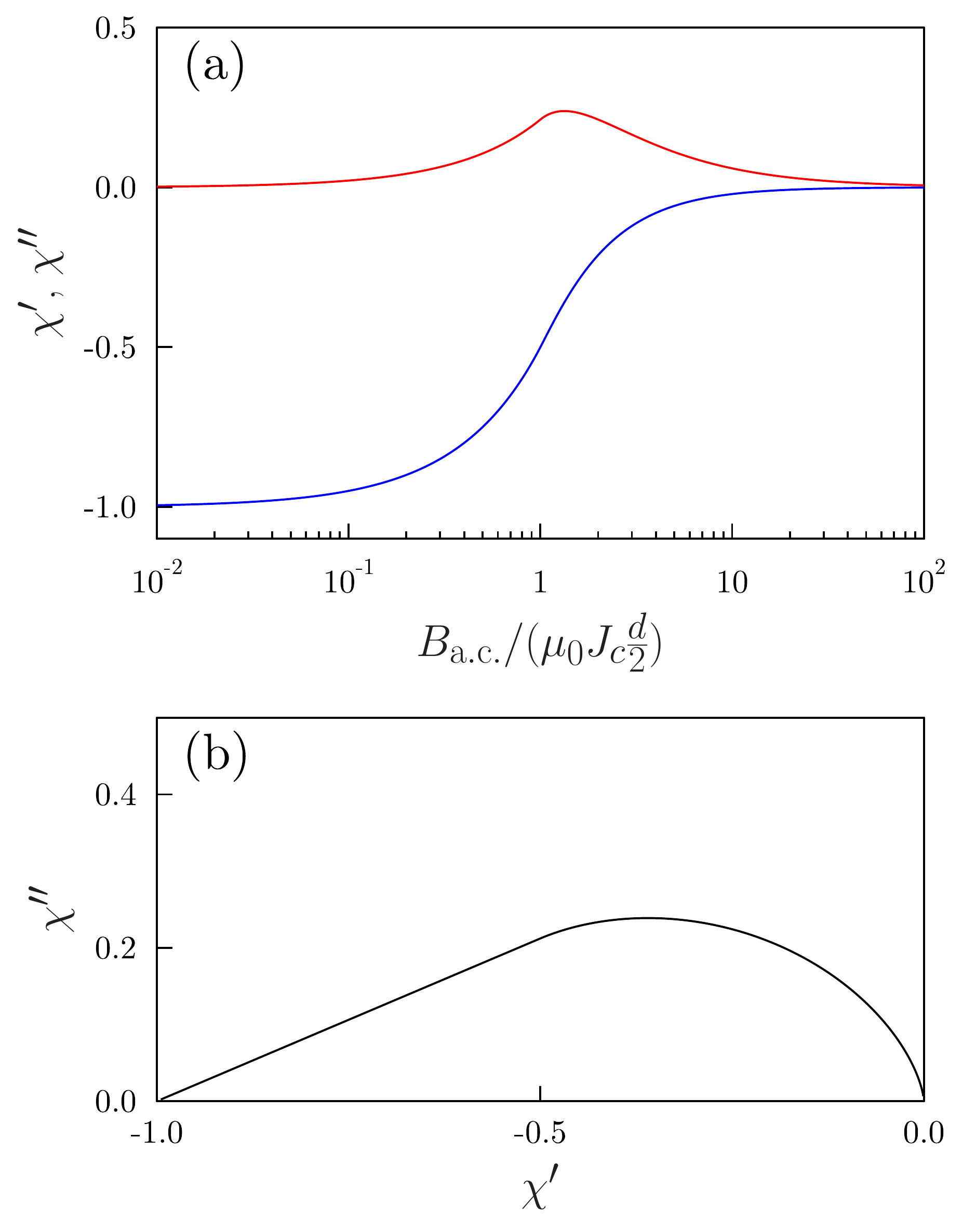}%
		\caption{(a) $\chi'$ and $\chi''$ and  (b) the Cole-Cole plot for Bean's critical state model \cite{Bean1964} for a slab of width $d$ (see~\cite{Ji1989}).\label{fig:criticalstate}}
\end{figure}

When flux lines are no longer weakly pinned the a.c.\ response  can be treated by using the notion of the critical state (introduced by Bean \cite{Bean1964}) in which the field gradient (related to the current density) takes its maximum value throughout the sample and the flux everywhere is on the point of slipping, though surface pinning can also be important \cite{Gomory1997,HandbookSupMat}.  The real and imaginary parts of $\chi$ can be calculated \cite{Ji1989} for Bean's model and the results are shown in figure~\ref{fig:criticalstate} for the slab geometry, demonstrating that the Cole-Cole plot can be studied as a function of the a.c.\ driving amplitude $B_{\rm a.c.}$.

In the critical state regime (where flux lines are no longer weakly pinned) the a.c.\ response becomes nonlinear.  Therefore, 
\begin{equation}
   { M(t) \over H_{a.c.}} = \sum_{n=1}^{\infty} \mbox{Im}(\chi_n {\rm e}^{i\omega t}),
\end{equation}
becomes a more appropriate description of the system where higher harmonics are explicitly included with the $n$th harmonic defined as $\chi_n=\chi_n^\prime-{\rm i}\chi_n^{\prime\prime}$ \cite{VanderBeek1993, Ishida1990, Nikolo2016, Qin1996,HandbookSupMat}.  A driving force of the form $H(t)=H_{\rm a.c.}\mbox{Im}({\rm e}^{i\omega t}) = H_{\rm a.c.}\sin\omega t$ has been assumed.  Here, the ``usual'' susceptibility $\chi^\prime - i\chi^{\prime\prime}$ is the first harmonic $\chi_1^\prime - i\chi_1^{\prime\prime}$.  In this formulation, the harmonics can be extracted from
\begin{eqnarray}
  \chi_n' &=& {1 \over \pi H_{\rm a.c.}} \int_0^{2\pi} M(t)\sin(n\omega t)\,{\rm d}(\omega t) \\
    \chi_n'' &=& -{1 \over \pi H_{\rm a.c.}} \int_0^{2\pi} M(t)\cos(n\omega t)\,{\rm d}(\omega t).
\end{eqnarray}
Other conventions can be used to define these harmonics (see for example the Appendix to \cite{Ishida1990} where a cosine a.c.\ drive field was assumed).  Measurement of higher harmonics in a.c.\ studies of superconductors has been used to test aspects of critical state models \cite{Ishida1990,Bean1964,Ji1989} as well as differentiate between flux creep models \cite{Adesso2004} and the dynamic behaviour of the vortex lattice in the liquid and glass regimes \cite{Polichetti2004}.  Cole-Cole plots of the third harmonics show distorted closed loops providing further data for which to test applicable models \cite{Adesso2004,Adesso2004b}.  The third harmonics measured in LaFeAsO$_{1-x}$F$_x$ were used to infer effects of grain  \cite{Polichetti2012}.  
The peak in $\chi^{\prime\prime}(T)$ can be associated with the irreversibility line of superconductors~\cite{Nikolo2016,HandbookSupMat,Malozemoff1988,Gomory1993}.
In the linear regime there is a frequency dependence, but no dependence on the amplitude of $B_{\rm a.c.}$; in the critical state, the reverse is true but the critical current  
must be taken into account \cite{Gomory1993}.

The thermal activation of flux line motion was first treated by
Anderson and Kim \cite{Anderson1962, Anderson1964} and models each pinning centre as a potential well of depth, $U$ \cite{Nikolo2016, Nikolo2015, Nikolo1993}.  $U$ depends on the spatial size of the pinned flux line (or bundle of flux lines) and the bulk critical field or critical current density \cite{Nikolo1994}.  The rate, $\nu$, at which a bundle may hop between pinning centres is given by

\begin{equation}
\nu = \nu_0\exp\left[\frac{-(U-W)}{k_{\rm B}T}\right]-\nu_0\exp\left[\frac{-(U+W)}{k_{\rm B}T}\right],\label{SuppHop}
\end{equation}
where $W$ is an adjustment to pinning potential height due to the Lorentz force acting on the pinned flux from the applied magnetic field and $\nu_0$ is a characteristic hopping rate \cite{Anderson1962,Anderson1964}.  The above equation represents forward and backward hopping between two wells which is either helped or hindered by the effect of the magnetic field.  It can be rewritten as

\begin{equation}
    \nu=2\nu_0\sinh(W/k_{\rm B}T)\exp(-U/k_{\rm B}T),
\end{equation}
where for $W\ll k_{\rm B}T$ it reduces to

\begin{equation}
    \nu=2\nu_0\frac{W}{k_{\rm B}T}\exp(-U/k_{\rm B}T),
\end{equation}
which looks like an Arrhenius law with a $1/T$-scaled prefactor \cite{Nikolo1993}.  The condition $W\ll k_{\rm B}T$ (thermally activated flux flow) corresponds to the regime of weakly pinned flux and can be observed at high temperatures and with near constant applied magnetic fields (i.e.\ $B_{\rm a.c.}\ll B_{\rm d.c.}$) such that the current density does not vary greatly \cite{Nikolo1993}.
The opposite situation ($W\gg k_{\rm B}T$) is the flux creep regime in which the superconductor is in the critical state with $J\approx J_C$ \cite{Nikolo1993}.  Here the backward hopping term of Eq.~\ref{SuppHop} can be neglected due to the large $W$ \cite{Nikolo2016, Nikolo2015, Nikolo1993} yielding a hopping rate 

\begin{equation}
\nu = \nu_0\exp\left[\frac{-(U-W)}{k_{\rm B}T}\right]
\end{equation}
that is also of the Arrhenius form.
The mechanism described above assumes an identical potential well at each pinning site, but there is no reason why all pinning sites should be identical and probably  a distribution of potential well heights is more realistic \cite{VanderBeek1993}.  Nevertheless, as for SMMs and ferromagnets, one can record $\chi^{\prime\prime}(T)$ at various frequencies in order to extract an effective activation energy (yielding $U$ or $U-W$, depending on the regime), as long as the amplitude of the d.c.\ and a.c.\ fields are appropriate for observing an Arrhenius dependence \cite{Nikolo2015}.
Moreover, the mutual repulsion of the vortices within a bundle of pinned vortices can also contribute to the breakdown of an Arrhenius type law at higher a.c.\ amplitudes \cite{Nikolo1994}.

\begin{figure} [tbp]
\centering
   \fbox{\bf Copyrighted figure -- please see published article}
	\caption{a.c.\ magnetic susceptibility of the superconductor (Ba$_{0.6}$K$_{0.4}$)Fe$_2$As$_2$ from reference~\cite{Nikolo2015}.  The upper and lower panels show the real and imaginary parts of the susceptibility with magnetic fields of $\mu_0H_{\rm d.c.}=0~\rm T$ and $\mu_0H_{\rm a.c.}=0.1~\rm mT$ focused on a temperature range around the frequency dependent portion of the data.  Inset plots show the entire temperature range.\label{SC_AC}}
\end{figure}

In a typical frequency dependent a.c.\ measurement of a type-II superconductor, one expects $\chi^{\prime}$ to show a drop to represent superconducting diamagnetic shielding while $\chi^{\prime\prime}$ should display a peak as temperature is lowered (representing the point at which the a.c.\ magnetic field penetrates the centre of the sample) \cite{Gomory1997,Nikolo2015, Gomory1993}.  Both of these features are frequency dependent and have been observed in various type-II superconductors such as (Ba$_{0.6}$K$_{0.4}$)Fe$_2$As$_2$ which can be seen in figure~\ref{SC_AC} \cite{Nikolo2015} and Na(Fe$_{0.94}$Ni$_{0.06}$)$_2$As$_2$ \cite{Nikolo2016}.

It is worth noting that a complicating factor in a.c.\ measurements of type-II superconductors is due to the sample itself.  Should the sample be polycrystalline (i.e.\ a powder or pellet) the diamagnetic drop of $\chi^\prime$ with decreasing temperature can be considerably broader with a lower temperature hump while $\chi^{\prime\prime}$ can be dominated by a broad peak in addition to the much smaller peak below $T_{\rm c}$ we have associated with flux motion.  
In essence this is due to the sample showing two critical temperatures: one associated with the bulk superconductor (referred to as an intrinsic peak) and one associated with coupling between grain boundaries (known as the lower coupling peak) \cite{Nikolo1994,Nikolo1989,Goldfarb1991,Goldfarb1987}.  This means that the full diagmagnetic shielding of $-1$ cannot be reached at temperatures between these two $\chi^{\prime\prime}$ peaks.  It has been found that this lower coupling peak is much more sensitive to the applied field than the intrinsic peak and in a high quality sample  the two peaks may coincide for low measuring fields \cite{Goldfarb1987}.

\section{Conclusion}

In the study of both magnetism and superconductivity the
measurement technique of a.c.\ magnetic susceptibility has a lot to offer.  We have reviewed its theoretical foundations and explored how it can be used in a wide range of materials, highlighting the similarities and differences between different types of experimental system.  Of course, no technique should be used in isolation and a.c.\ susceptibility is most powerfully used when performed in combination with a number of other techniques that can probe the structural, thermal, electrical and magnetic properties of a single sample.  However, its particular focus on low-frequency dynamics allows the elucidation of relatively sluggish processes that are missed by other techniques, processes such as the movement of domain walls or the slow relaxation in nanomagnets.  Though many dynamical effects in condensed matter physics are very rapid, occurring on time scales of order $h/E$ (where $E$ is an energy which could be $E_{\rm F}$ or $J$ or $k_{\rm B}T_{\rm c}$, giving times typically in the sub-picosecond regime), an important number of emergent properties give rise to variations on a dramatically slower timescale and a.c.\ susceptibility is well suited to catching them.

\section{Acknowledgements}
We would like to thank the following for useful discussions: Martin
Nikolo, Nathaniel Davies, Tom Lancaster, Dharmalingam Prabhakaran.  We are grateful to EPSRC (UK) for financial support.

\appendix

\setcounter{section}{0}
\section{Generalized Debye model}
\begin{figure} [tbp]
\centering
	\includegraphics[width = 1\columnwidth]{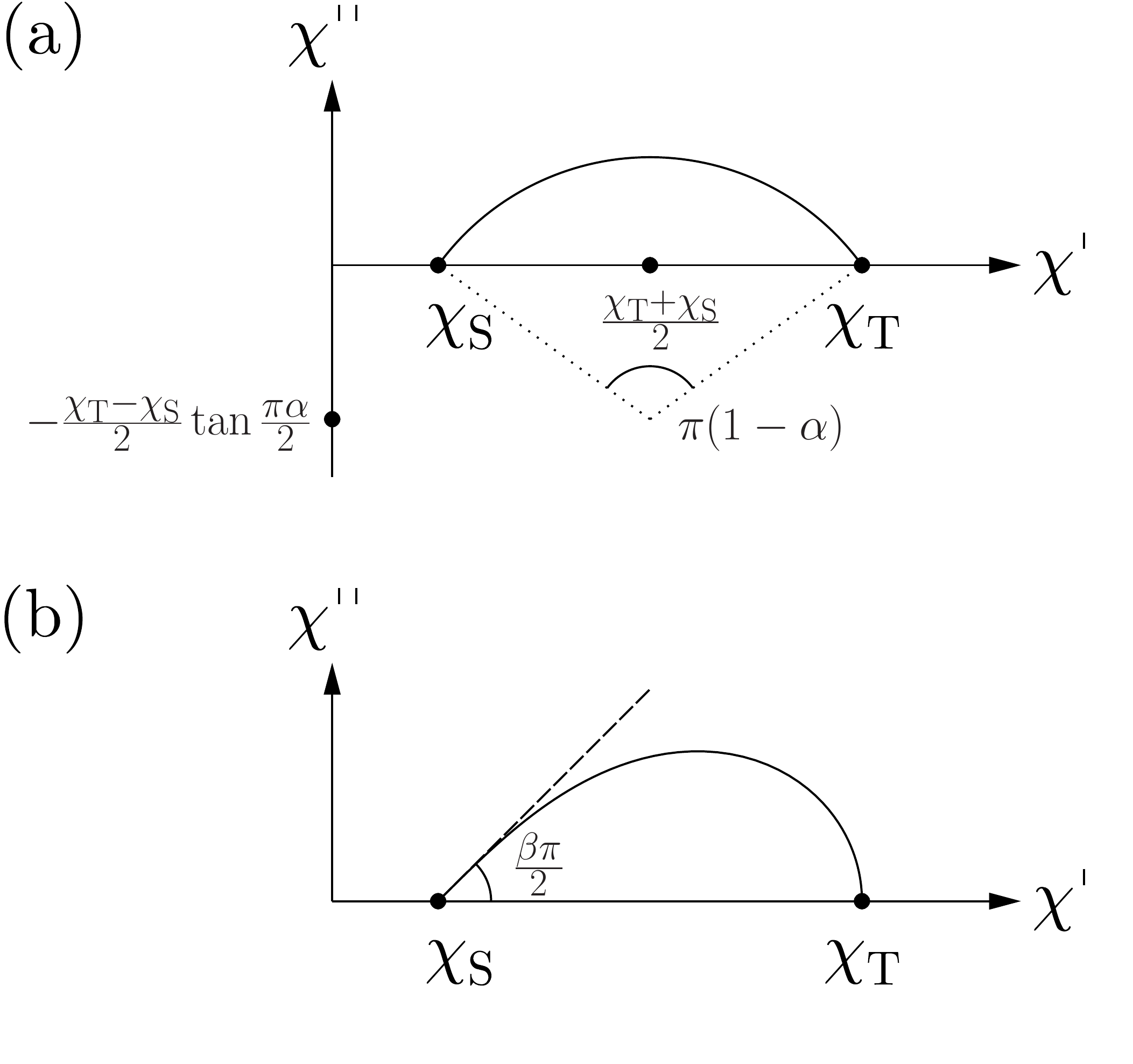}%
		\caption{(a) The generalized Debye model.  (b) The Cole-Davidson model\label{fig:response3}}
\end{figure}

The generalized Debye model is given by
\begin{equation}
  \chi(\omega)=\chi_{\rm S}+
  { \chi_{\rm T}-\chi_{\rm S} \over 1+({\rm i}\omega\tau)^{1-\alpha} }.
\end{equation}
Writing $\psi=(\omega\tau)^{1-\alpha}$
we have that $({\rm
  i}\omega\tau)^{1-\alpha} = \psi(\sin{\pi\alpha\over 2}+{\rm i}\cos{\pi\alpha\over 2})$ and hence the real and
imaginary parts of $\chi(\omega)$ can be written
\begin{eqnarray}
  \chi'(\omega) & = & \chi_{\rm S} +   { (\chi_{\rm T}-\chi_{\rm S})
    (1+\sin{\pi\alpha\over 2}\psi)
    \over 1+2\psi\sin{\pi\alpha\over 2}+\psi^2 } \\
    \chi''(\omega) & = &  { (\chi_{\rm T}-\chi_{\rm S})
    \cos{\pi\alpha\over 2}\psi
     \over 1+2\psi\sin{\pi\alpha\over 2}+\psi^2 . } 
\end{eqnarray}
An alternative form can be obtained by using the identities
\begin{eqnarray}
   {\psi+\psi^{-1} \over 2} & = & \cosh[(1-\alpha) \ln(\omega\tau) ]
   \\
   {\psi-\psi^{-1} \over 2} & = & \sinh[(1-\alpha) \ln(\omega\tau) ] ,
\end{eqnarray}
and hence
\begin{eqnarray}
  \chi'(\omega) & = & \chi_{\rm S} +   \frac{1}{2}(\chi_{\rm T}-\chi_{\rm S}) \times \nonumber \\
  & & 
  \left[
    1 - { \sinh[(1-\alpha) \ln(\omega\tau) ] \over
      \cosh[(1-\alpha) \ln(\omega\tau) ] + \sin{\pi\alpha\over 2} }
    \right] \label{eq:sinh} \\
    \chi''(\omega) & = & { \frac{1}{2}(\chi_{\rm T}-\chi_{\rm S})
      \cos{\pi\alpha\over 2} \over
      \cosh[(1-\alpha) \ln(\omega\tau) ] + \sin{\pi\alpha\over 2} }.
       \label{eq:coshh}
\end{eqnarray}
These equations can be rearranged to make the hyperbolic terms the
subject:
\begin{eqnarray}
   \sinh [(1-\alpha) \ln(\omega\tau) ] & = & {[ (\chi_{\rm T}+\chi_{\rm
     S})/2 - \chi' ]\cos\frac{\pi\alpha}{2} \over \chi'' } \\ 
   \cosh  [(1-\alpha) \ln(\omega\tau) ] & = &
   { (\chi_{\rm T}-\chi_{\rm S})\cos\frac{\pi\alpha}{2} \over
     2\chi''}
   - \sin\frac{\pi\alpha}{2}.
\end{eqnarray}
Then, using the identity $\cosh^2x-\sinh^2x=1$, we can show that
\begin{multline}
  \left( \chi' - \frac{\chi_{\rm T}+\chi_{\rm S} }{2} \right)^2
  + \chi''^2\sec^2{\pi\alpha\over 2} \\- \left(
  \chi''\tan\frac{\pi\alpha}{2}
  - \frac{\chi_{\rm T}-\chi_{\rm S}}{2} \right)^2 = 0,
\end{multline}
which can be simplified to
\begin{multline}
  \left( \chi' - \frac{\chi_{\rm T}+\chi_{\rm S} }{2} \right)^2
  + \left(\chi'' + \frac{\chi_{\rm T}-\chi_{\rm S}}{2}\tan\frac{\pi\alpha}{2} \right)^2
  = \\ \left( 
  \frac{\chi_{\rm T}-\chi_{\rm S}}{2}\sec\frac{\pi\alpha}{2}
  \right)^2 ,
  \label{eq:circle}
\end{multline}
which is the equation of a circle, with centre
\begin{equation}
  \left( \frac{\chi_{\rm T}+\chi_{\rm S} }{2}, -\frac{\chi_{\rm
      T}-\chi_{\rm S}}{2}\tan\frac{\pi\alpha}{2} \right)
\end{equation}
and radius $\frac{\chi_{\rm T}-\chi_{\rm S}}{2}\sec\frac{\pi\alpha}{2}$.
This provides the analytical form for the locus in the Cole-Cole plot,
as shown in figure~\ref{fig:response3}(a) and allows one to deduce that the opening angle
is $\pi(1-\alpha)$, as shown.  Equation~(\ref{eq:circle}) can be
rearranged to write $\chi''$ as a function of $\chi'$, yielding
\begin{multline}
  \chi'' = - \left(\frac{\chi_{\rm T}-\chi_{\rm S}}{2}\tan\frac{\pi\alpha}{2}\right)
  \\ \pm
  \sqrt{
   \left(\frac{\chi_{\rm T}-\chi_{\rm
       S}}{2}\tan\frac{\pi\alpha}{2}\right)^2
   + (\chi'-\chi_{\rm S})(\chi_{\rm T}-\chi')
  } , 
\end{multline}
which is equation~(\ref{ColeColeTest}).

\section{Cole-Davidson model}
A similar analysis can be carried out using the Cole-Davidson model
which is
\begin{equation}
  \chi(\omega)=\chi_{\rm S}+
  { \chi_{\rm T}-\chi_{\rm S} \over (1+{\rm i}\omega\tau)^{\beta} }.
\end{equation}
We can write the denominator as
\begin{equation}
(1+{\rm i}\omega\tau)^\beta = (1+\omega^2\tau^2)^{\beta/2} {\rm
    e}^{{\rm i}\beta\theta},
\end{equation}
where $\theta=\tan^{-1}(\omega\tau)$ and hence
\begin{equation}
  \chi(\omega)  =  \chi_{\rm S} +
  { \chi_{\rm T} - \chi_{\rm S} \over (1+\omega^2\tau^2)^{\beta/2} }
    {\rm e}^{-{\rm i}\beta\theta } .
\end{equation}
This gives the real and
imaginary parts of $\chi(\omega)$ as
\begin{eqnarray}
  \chi'(\omega) & = & \chi_{\rm S} +
  { \chi_{\rm T} - \chi_{\rm S} \over (1+\omega^2\tau^2)^{\beta/2} }
    \cos\beta\theta \\
  \chi''(\omega) & = & 
  { \chi_{\rm T} - \chi_{\rm S} \over (1+\omega^2\tau^2)^{\beta/2} }
    \sin\beta\theta .
\end{eqnarray}
Using $\cos\theta = (1+\omega^2\tau^2)^{-1/2}$ allows one to write
these equations in the alternate form
\begin{eqnarray}
  \chi'(\omega) & = & \chi_{\rm S} +
  ( \chi_{\rm T} - \chi_{\rm S} ) (\cos\theta)^\beta
    \cos\beta\theta \\
  \chi''(\omega) & = & 
  ( \chi_{\rm T} - \chi_{\rm S} ) (\cos\theta)^\beta
    \sin\beta\theta .
\end{eqnarray}
In the high frequency limit $\cos\theta \to 1/(\omega\tau)$ and $\theta\to\frac{\pi}{2}$
while $\cos\beta\theta \to \cos\frac{\beta\pi}{2}$.
This means that $\chi''\to (\chi'-\chi_{\rm S})
\tan\frac{\beta\pi}{2}$
and so the angle between the $\chi''(\chi')$ curve and the horizontal
axis approaches $\frac{\beta\pi}{2}$ in the Cole-Cole plot near
$\chi'=\chi_{\rm S}$ (see figure~\ref{fig:response3}(b)).  The condition for the maximum value of $\chi''$ can be obtained
by differentiation, yielding $\theta=\tan^{-1}(\omega\tau)=\frac{\pi}{2(\beta+1)}$.

\section{Havriliak-Negami model}
A similar derivation can be used to show that the Havriliak-Negami model which is
\begin{equation}
  \chi(\omega)=\chi_{\rm S}+
  { \chi_{\rm T}-\chi_{\rm S} \over (1+({\rm i}\omega\tau)^{1-\alpha})^{\beta} }
\end{equation}
leads to \cite{Havriliak1967,alvarez1991}
\begin{eqnarray}
  \chi'(\omega) & = & \chi_{\rm S} +
    {  ( \chi_{\rm T} - \chi_{\rm S} ) \cos\beta\theta \over \left[ 1+2(\omega\tau)^{1-\alpha}\sin{\pi\alpha\over 2} + (\omega\tau)^{2(1-\alpha)} \right] }\\
  \chi''(\omega) & = & 
    { ( \chi_{\rm T} - \chi_{\rm S} ) \sin\beta\theta \over \left[ 1+2(\omega\tau)^{1-\alpha}\sin{\pi\alpha\over 2} + (\omega\tau)^{2(1-\alpha)} \right] } ,
\end{eqnarray}
where
\begin{equation}
    \theta = \tan^{-1} \left[ {(\omega\tau)^{1-\alpha}\cos{\pi\alpha\over 2}\over 1+(\omega\tau)^{1-\alpha}\sin{\pi\alpha\over 2}}\right].
\end{equation}

\section*{References}
\bibliographystyle{unsrt}
\bibliography{ACSus-arXiv}

\end{document}